\documentclass[aps,prx,preprint,groupedaddress]{revtex4-1}

\usepackage[T1]{fontenc}
\usepackage[latin9]{inputenc}
\usepackage{float}
\usepackage{amsmath}
\usepackage{amssymb}
\usepackage{graphicx}
\usepackage{esint}
\usepackage{amsmath}

\begin{document}


\title{Electron emission by long and short wavelength lasers: essentials
for the design of plasmonic photocathodes}


\author{Ebrahim Forati}
\email[]{forati@ieee.org}
\affiliation{Applied Materials Inc.}

\author{Dan Sievenpiper}
\affiliation{University of California San Diego}

\date{\today}

\begin{abstract}
Theory of electron emission by metallic photocathodes under the exposure
of long wavelength lasers will be studied.  Energy of photons
in long wavelength lasers is less than the work function of the photocathode's
material, and can only emit electrons via tunneling through the potential
barrier. The optical resonance effect (e.g. plasmonic resonances) will be studied as an improvement
to the performance of photocathodes. This paper is intended to provide
self-sufficient materials to design optical resonant
surfaces (e.g. metasurfaces) for electron emission applications. 
\end{abstract}

\pacs{}
\keywords{Electron emission, photoemission, photoelectric, photocathode, plasmonic, metasurface}

\maketitle

\section{Introduction}

Electromagnetic interaction, one of the four known fundamental forces
in the universe, was proposed by Michael Faraday in 1820's \cite{faraday1834experimental},
theorized by James Clerk Maxwell in 1865 \cite{maxwell1881treatise},
and experimentally proved by Heinrich Hertz in 1888 \cite{hertz1893electric}. Since then, developed sources
of electromagnetic waves are based on moving (ideally resonating) electrons either in free space or inside
a material (e.g. semiconductors). In fact, less than four decades after the discovery of electromagnetism, in 1924, Louis de Broglie suggested that moving particles such as electrons behave as waves with frequencies
related to their momentums \cite{de1923waves}. Modern and traditional high power electromagnetic sources such as magnetrons and Klystrons
(developed during WWII by the Allied and the Axis Powers, respectively) are based on extracting electrons
into vacuum (electron emission), bunching them, modulating their speed,
and extracting their energy to an output electromagnetic wave.  In other words,
although transistors were invented more than half a century ago, modern
solid state microwave sources are still inferior to their electron beam-based
rivals at least in terms of the power level. A solid state source
is roughly limited to less than a MW peak power while an electron
beam-based source can provide GW ranges \cite{schamiloglu2004high}. This seems
obvious since semiconductors (and generally any material) impose their
limitations such as speed, bandgap, loss, etc. to the
designed device. As a matter of fact, residing electrons in vacuum, prior to their manipulation,
 is the extreme of relaxing most limitations of solid state devices. This signifies the importance and profundity of electron emission and its great potentials.   

Traditionally, thermal emission used to be the dominant method of electron
emission in engineering devices, which was followed by electric
field emission, photoemission, and even emission by another electron
beam (a.k.a. secondary emission). For the sake of self consistency,
we review these physics and their common formulations
in the next section, prior to discussing photoemission at long wavelength. The main contribution in this paper
is to formulate photoemission by photons with energies lower than
the photocathode's work function (e.g. infrared (IR) laser illuminating
gold photocathodes). Our motivation for this study is the potential
that we see in engineered resonant surfaces (a.k.a. metasurfaces)
for IR photocathode applications. Such photocathodes can be useful
at least for two purposes: 

a) IR photocathodes can be used to generate a cloud of free electrons
for microelectronic applications in order to circumvent semiconductors
limitations. For a semiconductor-free microelectronic device, IR is
much preferred over the usual ultraviolet (UV) light for a photocathode,
due to safety concerns. Moreover, metallic metasurfaces at IR can
leverage plasmoinc resonance effects to enhance the photon-electron
interaction significantly, which can reduce the light intensity requirement
down into the safe range. We reported an example of such devices in
\cite{forati2016photoemission}, where the continuous wave (CW) IR
radiation of a mW diode laser generated observable electron emission
using a properly designed plasmonic metasurface. 

b) metasurfaces are, in general, designed to manipulate/control the
phase front of the reflected wave from their surface \cite{sievenpiper2003two}.
If the metasurface itself is also the electron emitter, the correlation
between the incoming photons and emitted electrons may be maintained
to some extent. This may provide an opportunity to control the phase
front of the emitted electron beam (considering De Broglie wave picture
for electrons). Obviously, this possible application is not limited
to the IR range, but the ability to leverage plasmonic resonances
on noble metals makes the IR range more attractive. 

Nonetheless, there are stablished theories for UV photoemission as we will review later
in this paper.  Photocathodes have advanced applications such as in Linear accelerators
(to convert photon bunches into electron bunches), x-ray sources,
high energy colliders, and free electron lasers (FELs) where short
electron bunches are necessary. An FEL generates coherent light due
to the interaction of a train of electron bunches and light waves.
Important quantifying parameters of photocathodes are quantum efficiency
(QE), survivability, emission promptness, lifetime, and emittance, among which QE is extremely important since it is accessible experimentally. QE is defined as \cite{jensen2005time}
\begin{equation}
QE=\frac{\hbar\omega}{q}\left(\frac{J_{\mathrm{e}}}{I_{\mathrm{\lambda}}}\right)=1.2398\frac{J_{\mathrm{e}}\left(A/cm^{2}\right)}{I_{\mathrm{\lambda}}\left(W/cm^{2}\right)\lambda\left(nm\right)},
\end{equation}
in which q is the elementary charge, $\omega$ is the radial frequency, $\hbar$ is the reduced Plank's constant, $J_{\mathrm{e}}$ and $I_{\mathrm{\lambda}}$ are electric current and optical
intensity, respectively (units are given in the parenthesis). An example
of a photocathode requirements for FEL application is $0.1-1\,$A
current arranged in $0.1-1\,$ nC bunches in $10-50\,$ps pulses with
GHz repetition rates in applied fields of $10-50\,$MV/m \cite{jensen2006photoemission}.
The speed requirement rules out thermionic emitters for FELs since
they cannot switch current in picosecond time scales. In general,
an electron beam with a lower spatial divergence (emittance) and a
higher current (brightness) can provide shorter wavelength and more
powerful FELs. 

Both metals and semiconductors are popular for photocathodes. Metallic
photocathodes have more ruggedness, lower QE, and faster response
(which makes them suitable for pulse shaping). Metallic photocathodes
usually require UV light for emission, which normally
is considered as a restriction because UV is usually generated by
nonlinear conversion of 1064 nm Nd:YAG with low conversion efficiencies.
On the other hand, semiconductor photocathodes have higher QE, but
are chemically reactive and can be easily damaged by carbon dioxide
or water. They also have slower response time (larger than 40 ps)
compared to their metallic counterparts. Examples of semiconductor
photocathodes are direct bandgap p-type semiconductors including III-V
with cesium and oxidant, alkali antimonides, and alkali tellurides
\cite{jensen2006photoemission}. A usual solution for combining benefits
of rugged materials with high QE materials is to design dispenser
photocathodes, in which a low work function material such as Barium
diffuses out from a porous material with higher ruggedness (e.g. tungsten).
High vacuums (on the order of 1e-9 Torr) are usually needed for metallic
photocathodes, and the requirement becomes more strict for semiconductor
photocathodes.

Although approaches are similar, the physics of photoemission from
semiconductor and metallic photocathodes are different in their electron
distribution (e.g. bandgap) and transport (e.g. scattering mechanisms).
In this paper, we only consider metallic photocathodes although sum
of the discussions apply to semiconductor photocathodes as well. As
a result, Fermi-Dirac model for electrons distribution is used
throughout this paper. The arguments in this paper are valid for both
CW or pulsed laser excitations. 

\section{Thermal and electric field emission theories }

\begin{figure}
\includegraphics[width=3.5in]{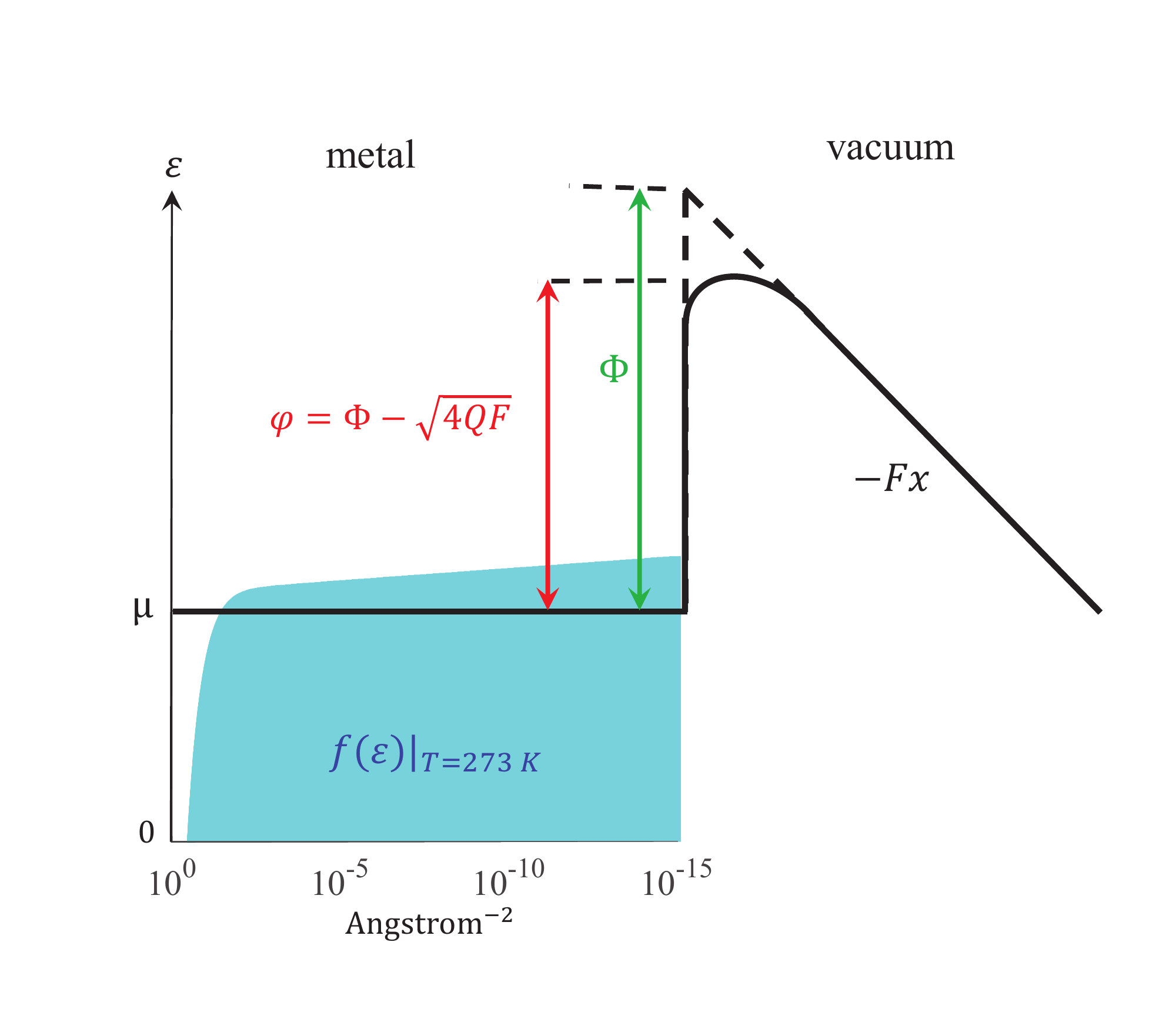}\caption{\label{fig: barrier}The potential barrier with Shottkey lowering
effect. The supply function of gold with $\mu=$5.53 eV at T=273 K
is also shown.}
\end{figure}

Four major identified methods of electron emission are thermal emission,
electric field emission, photoemission, and emission by other high
energy electrons (which is called secondary emission and is beyond
the subject of this paper). A photocathode (despite its name implication)
usually involves all three mechanisms. That is, usually a bias voltage
is applied to the photocathode, and usually the applied laser pulse
causes thermal heating of the photocathode. The canonical equations
governing thermal emission, field emission, and photoemission are
Richardson-Laue-Dushman, Fowler-Nordheim, and Fowler-Dubridge, respectively.
In order to start reviewing the three methods, consider the potential
diagram of a metal-vacuum interface as shown in Fig. \ref{fig: barrier}. 

The chemical potential of electrons in the metal and the work function
are identified by $\mu$ and $\Phi,$ respectively. If a static electric
field $\left(E_{\mathrm{static}}\right)$ is applied on the metal surface, the
work function will be reduced to \cite{dowell2009quantum,gadzuk1973field}

\begin{equation}
\phi=\Phi-\sqrt{4QF_{\mathrm{static}}}\label{eq:phi}
\end{equation}
 in which $F_{\mathrm{static}}=qE_{\mathrm{static}}$, 
and $Q=q^{2}/\left(16\pi\varepsilon_{0}\right)=0.35999$ eV$\,$nm.
The parameter $\phi$ is referred to as Shottkey reduced boundary
condition, and it includes the charge image effect of the metal on
the potential energy outside the metal (the same well-known image
theorem in electromagnetism). Since we only discuss metallic photocathodes
(with free electrons which are indeed Fermions), Fermi-Dirac distribution
is usually considered for electrons inside the metal and the number
density of electrons is given by \cite{jensen2017introduction}

\begin{equation}
\rho=\frac{2}{\left(2\pi\right)^{3}}\iiint\frac{dk}{1+exp\left(\beta_{\mathrm{T}}\left(\varepsilon-\mu\right)\right)},\label{eq:FD}
\end{equation}
where k is electron momentum and is related to its energy $\left(\varepsilon\right)$
by $\varepsilon=\hbar^{2}k^{2}/2m.$ The parameter $\beta_{\mathrm{T}}=1/k_{\mathrm{B}}T$
is sometimes called the thermal slope factor, in which $T$ is the
temperature in Kelvin unit and $k_{\mathrm{B}}$ is Boltzmann's constant. Equation
(\ref{eq:FD}) includes electrons with momentums in every direction,
however we are only interested in electrons with momentums normal
to the metal surface. It is straight forward to show that integration
of (\ref{eq:FD}) over parallel momentums to the metal surface leads
to \cite{jensen2017introduction}

\begin{equation}
f\left(\varepsilon\right)=\frac{m}{\pi\hbar^{2}\beta_{\mathrm{T}}}ln\left\{ 1+exp\left[\beta_{\mathrm{T}}\left(\mu-\varepsilon\right)\right]\right\} \label{eq:FD_2}
\end{equation}
which is called the supply function in electron emission theories.
In fact, (\ref{eq:FD_2}) is the number density of electrons with
momentums normal to the surface $\left(f\left(\varepsilon_{\bot}=\hbar^{2}k_{\bot}^{2}/2m\right)\right),$
which we dropped the subscript $\bot$ for convenience throughout
the rest of this paper. Figure \ref{fig: barrier} also includes the
supply function of gold with $\mu=5.53\,$eV at T=273$\,$K. 

The current density normal to the metal surface can be written
as \cite{jensen2017introduction}

\begin{equation}
\begin{aligned}
J=q\rho v=\frac{2q}{2\pi}\int\left(\frac{\hbar k}{m}\right)f\left(\varepsilon\right)T\left(\varepsilon\right)dk\\ 
\quad =\frac{q}{2\pi\hbar}\intop_{0}^{\infty}T\left(\varepsilon\right)f\left(\varepsilon\right)d\varepsilon,\label{eq:TF}
\end{aligned}
\end{equation}
in which $v=\hbar k/m$ is the speed of electrons, $T\left(\varepsilon\right)$
is the probability of an electron passing the barrier, and the factor
of 2 in the numerator accounts for the spin. 

In the event of a thermal emission, Richardson approximation is used,
where Heaviside step function is used for $T\left(\varepsilon\right)$ as

\begin{equation}
T\left(\varepsilon\right)=\begin{cases}
\begin{array}{c}
1\\
0
\end{array} & \begin{array}{c}
\varepsilon>\phi\\
\varepsilon<\phi
\end{array}\end{cases}.
\end{equation}

This leads to the well-known Richardson-Laue-Dushman equation for
thermal emission as \cite{jensen2006photoemission}

\begin{equation}
J_{\mathrm{RLD}}=A_{\mathrm{RLD}}T^{2}exp\left(-\beta_{\mathrm{T}}\phi\right),\label{eq:RLD}
\end{equation}
where $A_{\mathrm{RLD}}$ is a constant given by $A_{\mathrm{RLD}}=qmk_{\mathrm{B}}^{2}/\left(2\pi^{2}\hbar^{3}\right)=120.17\;\mathrm{A/cm^{2}\,K^{2}}$.
Richardson approximation assumes only electrons with energies higher
than the work function can escape, and neglects the tunneling emission.

Note that in the absence of large applied static fields, for average
laser illuminations, the thermally-driven emission is usually not significant.
For example, as calculations are done in \cite{jensen2006photoemission},
for a laser intensity of 28$\,\mathrm{MW/cm^{2}}$, a field of 8$\,\mathrm{MV/m}$, a
Gaussian laser pulse with a 2.7 ns time constant, and a wavelength
of 800 nm, the thermal current is calculated to be 3.8 \% of the photoemission
current for a coated tungsten surface with the work function of 1.9$\,\mathrm{eV}.$ Nonetheless, heating has a great impact on scattering rates
and emission probability of electrons, and is important
to be included in models. Typically, temperatures above 1000 K are needed
for considerable thermal emissions. 

In the event of electric field emission, electrons do not have enough
energy to travel over the barrier. Instead, $T\left(\varepsilon\right)$
in (\ref{eq:TF}) should be obtained by finding the tunneling probability
of electrons through the barrier. In 1928, Fowler and Nordheim published
their cold field emission (CFE) paper, first based on a triangular
potential barrier and then based on Shottkey reduced barrier using
JWKB approximation \cite{fowler1928electron}. In 1956, Murphy and
Good derived the equation again \cite{murphy1956thermionic}, and
in 2006 Forbes added a very useful analytical approximation to obtain
the standard FN equation as \cite{deane2008formal,forbes2001new,jensen2003electron}

\begin{equation}
J_{\mathrm{FN}}=\frac{aF_{\mathrm{static}}^{2}}{\Phi q^{2}t^{2}}exp\left(\frac{-\nu bq\Phi^{1.5}}{F_{\mathrm{static}}}\right)\label{eq:FN}
\end{equation}
where $a=q^{3}/\left(8\pi h\right)$ and $b=8\pi\sqrt{2m}/\left(3qh\right)$
are first and second FN constants, respectively. Both $t$ and $\nu$
can be fount analytically using Forbe's approximation as \cite{forbes2006simple}

\begin{equation}
v=\left(1-y^{2}\right)+\frac{1}{3}y^{2}ln\left(y\right),\label{eq:v}
\end{equation}

\begin{equation}
t=1+\frac{1}{9}y^{2}\left(1-ln\left(y\right)\right),\label{eq:t}
\end{equation}
both of which depend on Nordheim's parameter as 
\begin{equation}
y=\sqrt{\frac{q^{2}F_{\mathrm{static}}}{4\pi\varepsilon_{0}\Phi^{2}}}.
\end{equation}

\begin{figure}
\begin{minipage}[t]{0.45\columnwidth}%
\includegraphics[width=\textwidth]{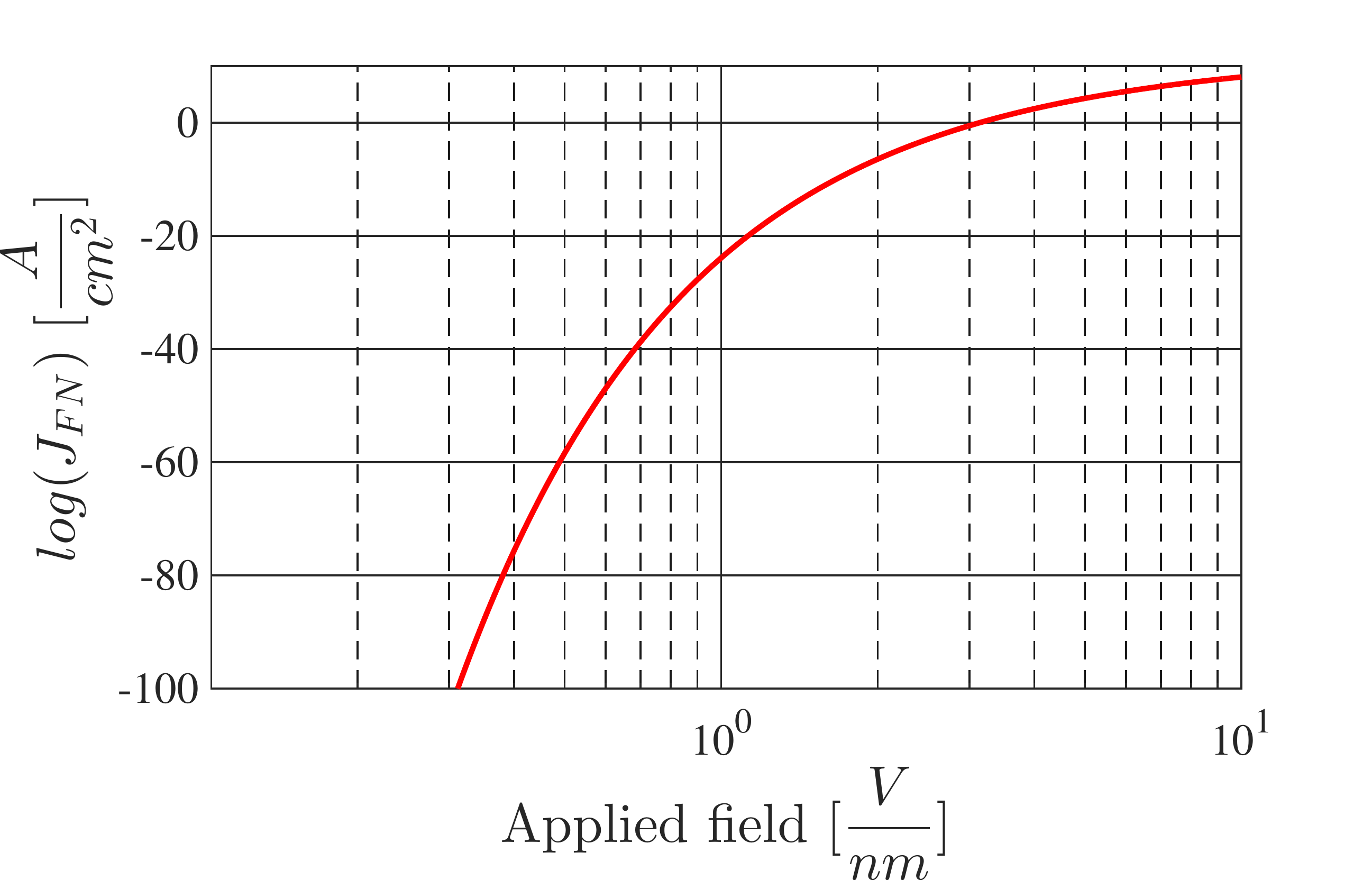}

(a)%
\end{minipage}%
\begin{minipage}[t]{0.45\columnwidth}%
\includegraphics[width=\textwidth]{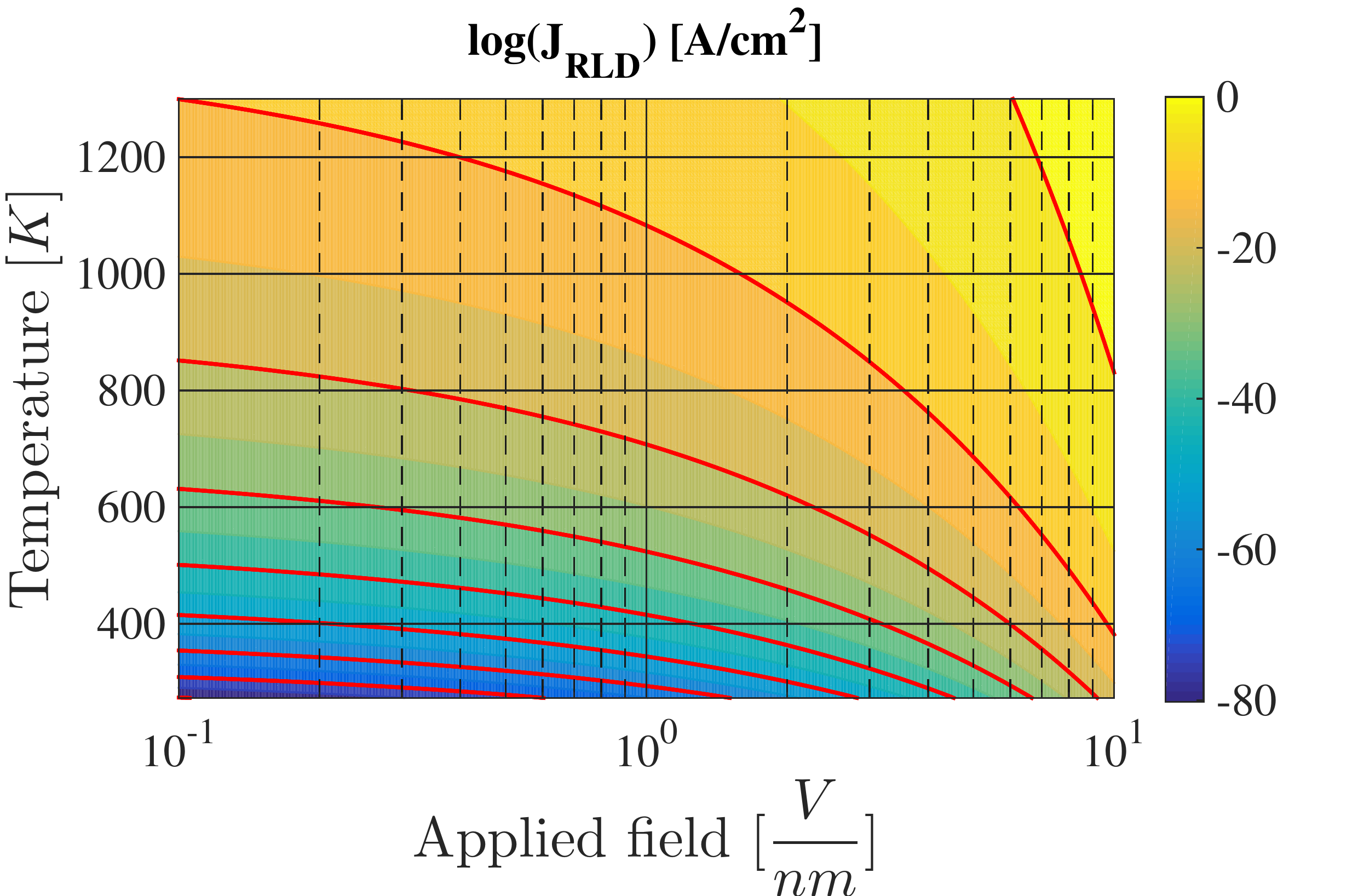}

(b)%
\end{minipage}

\caption{a) \label{fig:FN_RLD}FN emission (temperature independent), b) thermal
emission (red lines indicate isocurrent curves starting from 0 with
steps of -10$\,\mathrm{A/cm^{2}}$). In both calculations, gold's Fermi
level of 5.53 eV and work function of 5.1 eV were used. }

\end{figure}

It is useful to compare the contribution of these emissions as a function
of the applied electric field and temperature, as shown in Fig. \ref{fig:FN_RLD}.
$J_{\mathrm{FN}}$ is plotted only as a function of the applied static field
since it is temperature independent. 

In calculations of tunneling probability through a barrier, it is
common to define a Gamow factor, $\theta\left(\varepsilon\right),$
so that $T\left(\varepsilon\right)\propto e^{-\theta\left(\varepsilon\right)}.$  The Gamow factor carries the barrier shape information and is defined
as (WKB formulation) 

\begin{equation}
\theta\left(\varepsilon\right)=2\intop_{x_{-}}^{x_{+}}\sqrt{\frac{2m}{h^{2}}\left(V\left(x\right)-\varepsilon\right)}dx,\label{eq:Gamow}
\end{equation}
where $V\left(x_{\pm}\right)=\varepsilon.$ In fact, $\theta\left(\varepsilon\right)$ is
defined as the area under the curve of the potential barrier ($V\left(x\right)$) between $x_{-}$ and $x_{+}$.
A known example of defining $T\left(\varepsilon\right)$ based on
the Gamow factor is given by $T\left(\varepsilon\right)=1/\left(1+e^{\theta\left(\varepsilon\right)}\right)$
\cite{kemble1935contribution}. There are two more parameters which are
useful to review before moving to the next section. The value of
$\varepsilon$ which maximizes the integrand in (\ref{eq:TF}) is
of special importance and is usually specified by $\varepsilon_{\mathrm{m}}.$
The field slope function, $\beta_{\mathrm{F}}$, and the slope factors ratio,
n, are also defined as

\begin{equation}
\beta_{\mathrm{F}}=-\frac{\partial}{\partial\varepsilon}\theta\left(\varepsilon=\varepsilon_{\mathrm{m}}\right),
\end{equation}

\begin{equation}
n=\beta_{\mathrm{T}}/\beta_{\mathrm{F}}.
\end{equation}

It is also important to know the limitation of using (\ref{eq:FN})
which is due to the zero temperature approximation of the supply function
in finding the tunneling probability. This approximation is valid for low,
including room, temperatures as the tunneling probability is dominated
by the contribution of electrons around the chemical potential level. 

So far we have reviewed thermal and field emissions separately. If
the applied filed is negligible or the temperature is low, one can
use (\ref{eq:RLD}) or (\ref{eq:FN}), respectively. However, often
these two physics are involved simultaneously which requires the use of
the general thermal field (GTF) emission theory (readers can find detailed
discussions about GTF in \cite{jensen2006general}). Based on GTF,
we may consider three operating regimes for an electron emitter: thermal,
field, and the transition regime where both thermal and field emissions
become comparable. 

In the next section, for engineering purposes and without going into
equations derivings, we summarize the steps to estimate thermal/field
electron emission when both the temperature and the applied filed
are known. 

\section{The general thermal-field emission model (with approximation)}

Several parameters of electric field emission such as scattering rates and electrons distribution are temperature dependent. Besides, electrons departure due to the electric field emission may change the emitter's temperature by extracting its energy. In fact, temperature is the key
parameter which couples field and thermal emissions together  in the general thermal-field emission theory. Details of
this theory is discussed in \cite{jensen2006general,jensen2007general},
and is briefly summarized in the appendix. Here, we use a critical
approximations which simplifies the formulation and provides an appropriate
estimate of the thermal and field emitted electrons from metallic
photocathodes. The approximation is based on the assumption that emitted
electrons do not change the photocathode's temperature. This is valid
if either the photocathode's operating temperature is low, or quantum
yield (QY) of the photocathode is small. QY of a photocathode is the
ratio of the emitted electrons to the penetrated photons (QY differs
from QE by a factor of photon's reflectivity from the photocathode's
surface). Despite the fact that metals reflectivity is very high, their QY  is a small number because of their  very small QE (less than 1\%).  To be more specific, the highest
experimental QE of metals is about 0.22\% and is produced by a laser
excitation of 0.6$\,\mathrm{mJ/cm^{2}}$ and under a 50$\,\mathrm{MV/m}$ static field.
Assuming gold as the photocathode (with the permittivity of  $\varepsilon_{\mathrm{r}}=-22-1.5i$
at 785 nm), the reflection coefficient of an IR wave is around 98.62\%
at normal incidence. Simple calculation shows that about 15\% of penetrated photons
are contributing to direct electron emission. We simply disregard
the cooling effect of these emitted electrons and assume that penetrated
photons are entirely absorbed by phonons in the metal, leading to
a temperature rise. Note that 0.6$\,\mathrm{mJ/cm^{2}}$ laser intensity is
very high, and QE of usual photocathodes is considerably smaller than
0.22\%. Nonetheless, there is always the choice of referring to the
appendix to perform more accurate and more complex calculations. 

In this low QY regime, total field-thermal emission contributions
of a photocathode (or any emitter) can be obtained using the following
steps (with the aid of commercially available electromagnetic solvers): 

1) The first step is to obtain the photocathode's approximate temperature
using electromagnetic and heat transfer calculations. The assumption
here is that electron and lattice temperatures are equal, and electron
emission does not affect the temperature. A simple method is to 
perform a full-wave electromagnetic simulation and find the absorbed
laser power by the photocathode. Then, the absorbed laser power can
be considered as the heat source in the heat transfer calculations.
This can be done analytically, or by a simulation solver along with
the appropriate convection, conduction, or radiation boundaries. Either
of the calculations should provide us the temperature distribution
on the photocathode. If the computational power allows us, a better
approach is to perform a coupled electromagnetic and heat transfer
simulation (e.g. using Multiphysics simulation is COMSOL). As a check
point, from the maximum temperature on the photocathode, we may validate
the approximation (i.e., to confirm the temperature is not too
high, e.g. >1000 K). We may consider the maximum calculated temperature
in the structure as the operating temperature, $T_{\mathrm{o}}.$ Note that
the maximum temperature obtained by this method is always larger than
the actual temperature since a fraction of the absorbed photons contribute
to direct photoemission. 

2) The second step is to determine the dominant operating regime of
the photocathode, which can be thermal, field, or the transition regime
(where thermal and field contributions are comparable). After finding
Shottkey reduced work function ($\phi$), and Fowler parameter ($y$),
thermal coefficients

\begin{equation}
B_{\mathrm{q}}=\frac{\pi}{\hbar}\phi\sqrt{2m}\left(\frac{Q}{F_{\mathrm{static}}^{3}}\right)^{1/4};\qquad C_{\mathrm{q}}=-\frac{B_{\mathrm{q}}}{\phi},
\end{equation}
and Fowler-Nordheim coefficients 

\begin{equation}
C_{\mathrm{FN}}=-\frac{2}{\hbar F}\sqrt{2m\Phi}t\left(y\right);\qquad B_{\mathrm{FN}}=\frac{4}{3\hbar F}\sqrt{2m\Phi^{3}}v\left(y\right),
\end{equation}

\begin{equation}
C_{\mathrm{n}}=-\frac{3B_{\mathrm{FN}}+\left(2C_{\mathrm{FN}}+C_{\mathrm{q}}\right)\phi}{\phi^{2}};\qquad D_{\mathrm{n}}=\frac{2B_{\mathrm{FN}}+\left(C_{\mathrm{FN}}+C_{\mathrm{q}}\right)\phi}{\phi^{3}}
\end{equation}
should be found, in which Forbes approximation are applicable to find
$v$ and $t$ as given by (\ref{eq:v})
and (\ref{eq:t}). Then, the two threshold temperatures $T_{\mathrm{min}}$
and $T_{\mathrm{max}}$ 

\begin{equation}
T_{\mathrm{min}}=-\frac{1}{k_{\mathrm{B}}C_{\mathrm{FN}}};\qquad T_{\mathrm{max}}=-\frac{1}{k_{\mathrm{B}}C_{\mathrm{q}}}
\end{equation}
determine the operating regime of the photocathode. Thermal and field
regimes are identified by $T_{\mathrm{o}}>T_{\mathrm{max}}$ and $T_{\mathrm{o}}<T_{\mathrm{min}}$, respectively.
The transition regime is identified by $T_{\mathrm{min}}<T_{\mathrm{o}}<T_{\mathrm{max}}.$ Detailed
explanations regarding these regimes and how the formulations are
derived are explained in \cite{jensen2007general}. 

4) The total current due to the thermal-field emission can be obtained
from \cite{jensen2017introduction}

\begin{equation}
J_{\mathrm{GTF}}=A_{\mathrm{RLD}}\left(k_{\mathrm{B}}\beta_{\mathrm{T}}\right)^{-2}N\left(n,s\right),
\end{equation}

\begin{equation}
N\left(n,s\right)=s_{\mathrm{ng}}+s_{\mathrm{n}}e^{\mathrm{-ns}}+n^{2}s_{\mathrm{d}}e^{\mathrm{-s}}.
\end{equation}

\begin{equation}
s_{\mathrm{d}}=-n^{-2}\left(0.10593434n^{-2}+0.35506593\right)-1,
\end{equation}

\begin{equation}
s_{\mathrm{n}}=-n^{2}\left(0.10593434n^{2}+0.35506593\right)-1,
\end{equation}

\begin{equation}
s_{\mathrm{ng}}=\begin{cases}
\begin{array}{c}
-2\left[e^{\mathrm{-ns}}-n^{4}e^{\mathrm{-s}}\right]\left[n^{2}-1\right]^{-1}\\
e^{\mathrm{-s}}\left[24s^{3}\left(s+4\right)-z^{3}\left(s^{4}-24\right)+4sz^{2}\left(s^{3}+24\right)-12s^{2}z\left(s^{2}-12\right)\right]\left[12s^{2}\left(2s+z\right)\right]^{-1}
\end{array} & \begin{array}{c}
\left|z\right|>0.1\\
\left|z\right|<0.1
\end{array},\end{cases}
\end{equation}
where $z=s\left(n-1\right),$ and parameters n and s depend on the operating regime (identified by $T_{\mathrm{o}}$),
and are given in Table I.
\begin{center}
Table 1: s and n for different operating regimes. 
\par\end{center}

\begin{center}
\begin{tabular}{|c|c|c|}
\hline 
 & s & n\tabularnewline
\hline 
\hline 
Thermal $\left(n<1\right)$ & $B_{\mathrm{q}}$ & $\phi\beta_{\mathrm{T}}/B_{\mathrm{q}}$\tabularnewline
\hline 
Field $\left(n>1\right)$ & $B_{\mathrm{FN}}$ & $-\beta_{\mathrm{T}}/C_{\mathrm{FN}}$\tabularnewline
\hline 
Transition $n=1$ & Eq. ((\ref{eq:s})) & $1$\tabularnewline
\hline 
\end{tabular}
\par\end{center}

In Table I, for the transition regime $\left(n=1\right)$, 

\begin{equation}
s=B_{\mathrm{FN}}-C_{\mathrm{n}}\left(\varepsilon_{\mathrm{m}}-\mu\right)^{2}-2D_{\mathrm{n}}\left(\varepsilon_{\mathrm{m}}-\mu\right)^{3},\label{eq:s}
\end{equation}

\begin{equation}
\varepsilon_{\mathrm{m}}=\mu-\frac{C_{\mathrm{n}}}{3D_{\mathrm{n}}}+0.5\sqrt{\left(\frac{2C_{\mathrm{n}}}{3D_{\mathrm{n}}}\right)^{2}-4\frac{\beta_{\mathrm{T}}+C_{\mathrm{FN}}}{3D_{\mathrm{n}}}}.
\end{equation}

\begin{figure}
\begin{minipage}[t]{0.45\columnwidth}%
\includegraphics[width=3in]{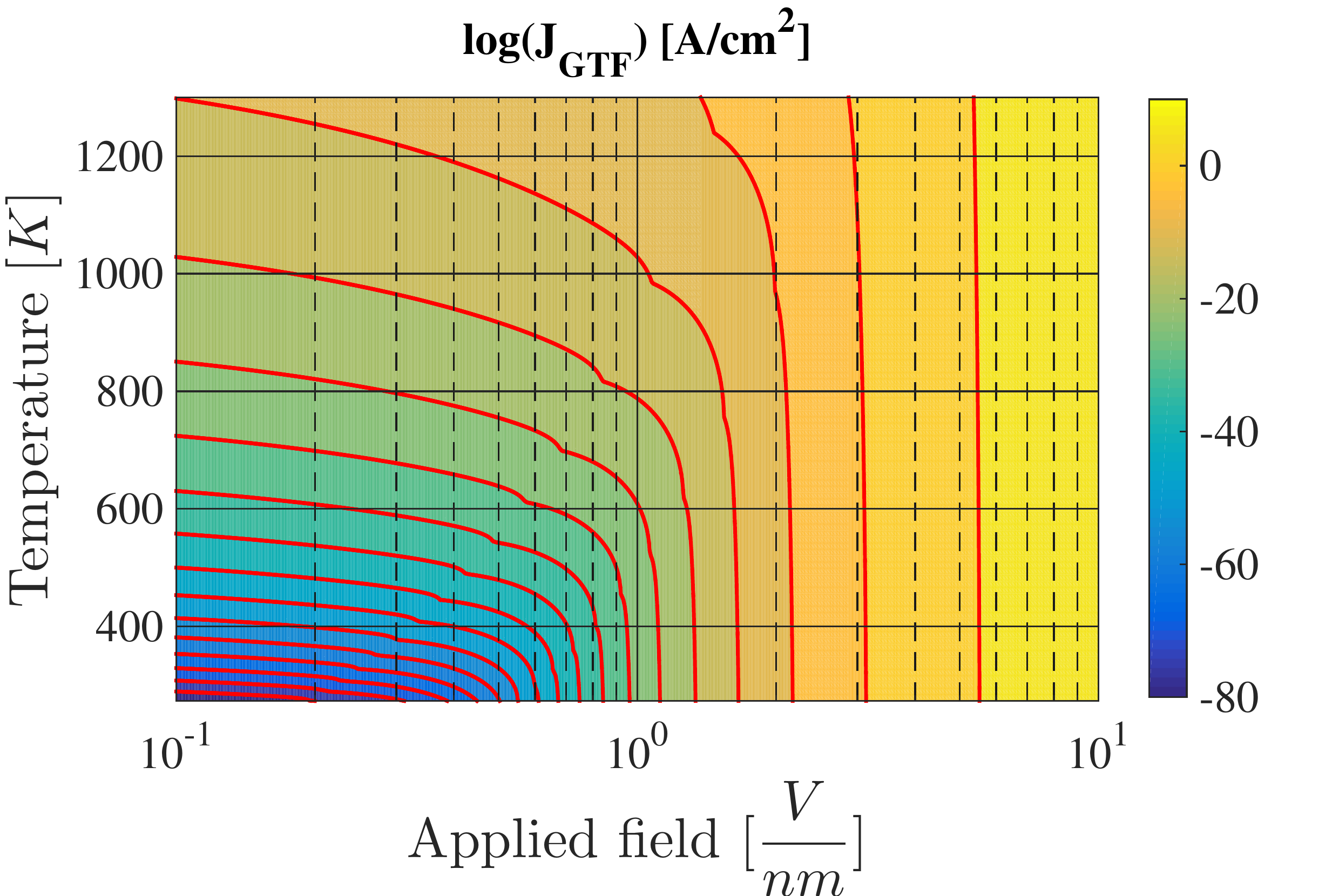}

(a)%
\end{minipage}%
\begin{minipage}[t]{0.45\columnwidth}%
\includegraphics[width=3in]{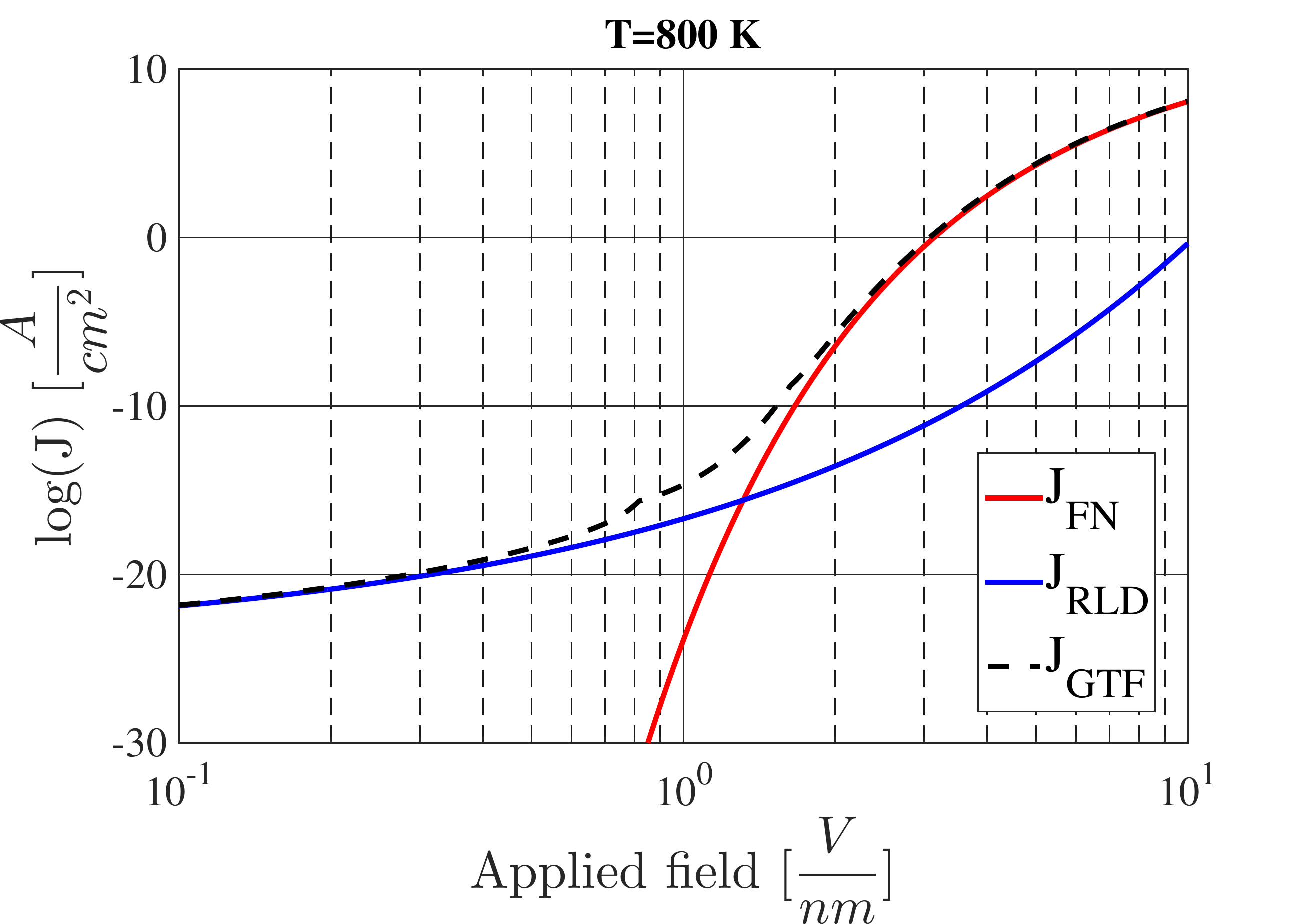}

(b)%
\end{minipage}

\caption{a)\label{fig:J_GTF} $J_{\mathrm{GTF}}$ as a function of the temperature and the
applied field (red lines indicate isocurrent curves starting from
5$\,\mathrm{A/cm^{2}}$ with steps of -5$\,\mathrm{A/cm^{2}}$), b) comparison among $J_{\mathrm{FN}}$,
$J_{\mathrm{RLD}}$, and $J_{\mathrm{GTF}}$ at T=800$\,$K. In all calculations, gold's
Fermi level of 5.53$\,$eV and work function of 5.1$\,$eV were used. }
\end{figure}

Before adding the photo-emitted contribution, it is useful to compare
different regimes as a function of the applied static field and the temperature.
Figure \ref{fig:J_GTF} shows $J_{\mathrm{GTF}}$ of a gold cathode.  By comparing $J_{\mathrm{GTF}}$, $J_{\mathrm{FN}}$, and $J_{\mathrm{RLD}}$ in Fig.  \ref{fig:J_GTF}(b), we can distinguish different operating 
regimes. The transition regime is where $0.5 \mathrm{\frac{V}{nm}}<F_{\mathrm{static}}<2 \mathrm{\frac{ V}{nm}}$.

\section{Photoemission by short wavelength lasers}

One of the most known models of photoemission is the three step model
of Spicer and Perrera-Gomez, developed in 1958 (a review of this model
is given in the appendix) \cite{spicer1993modern}. The three steps
in this model are a) optical absorption by the material, b) electron
transport inside the material, and c) electron emission (escape) from
the material surface. These three steps can be seen in Fowler-Dubridge
equation as 
\begin{equation}
J_{\mathrm{\lambda}}=I_{\mathrm{\lambda}}\left(\frac{q}{\hbar\omega}\right)\left(1-R\right)f_{\mathrm{\lambda}}P,\label{eq:J_lam-2}
\end{equation}
in which R is the reflection from the photocathode surface, $f_{\mathrm{\lambda}}$
is the probability of an excited electron to reach the surface without
experiencing any collision, and P is the probability of emission from
the metal surface. 

Having proper electromagnetic model of the photocathode's material
(e.g. permittivity model), it is fairly easy to calculate the photocathode
reflectivity (which leads to the photon absorption). In a one dimensional
geometry with normal laser incidence, reflectivity can be found from 

\begin{equation}
R=\left|\frac{1-\sqrt{\epsilon_{\mathrm{r}}}}{1+\sqrt{\epsilon_{\mathrm{r}}}}\right|,
\end{equation}
where $\epsilon_{\mathrm{r}}$ is the relative permittivity of the metal at
$\lambda.$ A more general equation for the reflectivity under oblique
incidence is given in the appendix. For complex geometries, numerical
Maxwell equation solvers can easily calculate the reflectivity or
the absorption of the photocathode. 

The second step is electrons travel to the surface, which usually
initiates by an applied static field drifting electrons to the surface.
On their path, these electrons experience electron-electron and electron-phonon
scatterings, as well as change of momentum directions. In most photoemission
models, the assumption is that a single collision extracts most of
the electron's energy and prevents it from contributing to direct
photoemission. Often, the scattering rate is replaced by the mean
free path of an electron, which is the collision-free travel range
of an electron in the metal. In metals, electron-electron scattering
is dominant and electrons mean free path can be obtained from \cite{dowell2006situ}

\begin{equation}
\lambda_{\mathrm{e-e}}=\frac{2\lambda_{\mathrm{m}}\varepsilon_{\mathrm{m}}^{3/2}}{\hbar\omega\sqrt{\phi}\left(1+\sqrt{\frac{\phi}{\hbar\omega}}\right)},
\end{equation}
where $\lambda_{\mathrm{m}}$ and $\varepsilon_{\mathrm{m}}$ are experimentally measured
energy and mean free path values which are available in \cite{dowell2006situ}.
$\delta$ is the skin depth which depends on the imaginary part of
the permittivity, $k$, as $\delta=\frac{\lambda}{4\pi k}$. Then, 

\begin{equation}
f_{\mathrm{\lambda}}=\intop_{0}^{\infty}\frac{1}{\delta}e^{-x\left(1/\delta+1/\lambda_{\mathrm{e-e}}\right)}dx=\left(1+\frac{\delta\left(\omega\right)}{\lambda_{\mathrm{e-e}}\left(\omega\right)}\right)^{-1}
\end{equation}

A more accurate equation for $f_{\mathrm{\lambda}}$ is to use (\ref{eq:f_lambda-1})
in the appendix, however a good approximation, which we will use throughout
this paper is 

\begin{equation}
f_{\mathrm{\lambda}}=0.5G\left(\frac{\delta m}{\hbar k_{\mathrm{v}}\tau}\right),\label{eq:F_lam}
\end{equation}

\begin{equation}
G\left(x\right)=\left\{ \begin{array}{cc}
1-\frac{2cos^{-1}\left(1/x\right)}{\pi sin\left(cos^{-1}\left(1/x\right)\right)} & x>1\\
1-\frac{1}{\pi}cot\left(cos^{-1}\left(x\right)\right)ln\left[\frac{1+sin\left(cos^{-1}\left(x\right)\right)}{1-sin\left(cos^{-1}\left(x\right)\right)}\right] & x<1
\end{array},\right.
\end{equation}
tin which $\hbar k_{\mathrm{v}}=\sqrt{2m\left(\mu+\phi\right)}$, $\tau$ is
the scattering rate (its value for gold at 537 K is 11.85 fs), and
$\delta$ is the skin depth. 

The third step is the emission of electrons from the metal surface.
The important effects in this step are Schottky effect, the abrupt
change in the electron angle across the metal-vacuum transition, and
the scattering (reflection) from the surface. In many models including
Fowler-Dubridge model, quantum effects such as tunneling and multiphoton
absorption are neglected, and only electron transport over the potential
barrier is considered. This is sometimes called the thermionic approximation,
and it should not be confused with the emission due to the thermal
effect. Based on Fowler-Dubridge mode, excited electrons can emit
only if their energy is higher than the work function. Therefore,
the probability of electron emission at the surface is 

\begin{equation}
P=\frac{\intop_{\mu+\phi-\hbar\omega}^{\infty}f\left(E\right)dE}{\intop_{0}^{\infty}f\left(E\right)dE}.\label{eq:P-2}
\end{equation}
Considering Fermi-Dirac distribution for electrons leads to 

\begin{equation}
P=\frac{U\left[\frac{1}{k_{\mathrm{B}}T}\left(\hbar\omega-\phi\right)\right]}{U\left(\frac{1}{k_{\mathrm{B}}T}\mu\right)},\label{eq:P}
\end{equation}
where $U$ is the Fowler function defined as

\begin{equation}
U\left(x\right)=\intop_{-\infty}^{\mathrm{x}}ln\left(1+e^{\mathrm{y}}\right)dy.\label{eq:U-1}
\end{equation}

There are several approximations for (\ref{eq:U-1}) such as \cite{jensen2005time}

\begin{equation}
U\left(x\right)=\begin{cases}
\begin{array}{c}
e^{\mathrm{x}}\left(1-0.17753e^{0.72843x}\right)\\
\frac{x^{2}}{2}+\frac{\pi^{2}}{6}-e^{\mathrm{-x}}\left(1-0.17753e^{\mathrm{-0.72843x}}\right)
\end{array} & \begin{array}{c}
x<0\\
x\geqslant0
\end{array}\end{cases}.
\end{equation}

There are also several approximations reported for (\ref{eq:P}) such
as the simple approximation \cite{dowell2006situ}

\begin{equation}
P=\frac{\left(\mu+\hbar\omega\right)}{2\hbar\omega}\left[1+\frac{\mu+\phi}{\mu+\hbar\omega}-2\sqrt{\frac{\mu+\phi}{\mu+\hbar\omega}}\right],
\end{equation}

and the more accurate equation \cite{jensen2007photoemission}

\begin{equation}
P=\left\{ \frac{3\left(\hbar\omega-\phi\right)^{2}+\beta^{2}\pi^{2}\left[1+n^{2}\right]}{6\hbar\omega\left(2\mu-\hbar\omega\right)}\right\} ,\label{eq:photoemission-1}
\end{equation}

in which 

\begin{equation}
n=\frac{1}{\pi k_{\mathrm{B}}T}\sqrt{\frac{\hbar^{2}}{2m}}\left(\frac{F_{\mathrm{static}}^{3}}{Q}\right)^{1/4}.
\end{equation}

A crucial point in all of these equations is the assumption that $\hbar\omega$
is larger than $\phi$, which is true in most photocathode designs
(e.g. UV photocathodes). However, (\ref{eq:P}) does not apply to
long wavelength photocathodes (e.g. IR) where there are potential
applications as discussed in the introduction. 

Note that $J_{\mathrm{\lambda}}$ is the emission by direct photo-excitation,
and is different from the emission stimulated by the laser heating.
For example, as calculated in \cite{jensen2006photoemission}, for a laser intensity
of 28$\,\mathrm{MW/cm^{2}}$ with a Gaussian pulse with 2.7$\,$ns time constant and a wavelength of 800$\,$nm, along with a static field of 8$\,\mathrm{MV/m}$  at room temperature, the thermal illumination-induced current is  3.8\% of the total photoemission current for a coated
tungsten surface with work function of 1.9$\,$eV. 

\section{Photoemission by long wavelength lasers}

The three step model (i.e. using (\ref{eq:J_lam-2})) applies to photoemission
by long wavelength lasers as well. However, since excited electrons
do not have enough energy to overpass the barrier, (\ref{eq:P-2})
needs to be generalized to include tunneling through the barrier.
In general, P as the emission probability of an electron (consecutive
to a photon absorption) at the photocathode's surface can be written
as 

\begin{equation}
P=\frac{\intop_{0}^{\infty}T\left(\varepsilon+\hbar\omega\right)f\left(\varepsilon\right)d\varepsilon}{\intop_{0}^{\infty}f\left(\varepsilon\right)d\varepsilon}.\label{eq:P-3}
\end{equation}
Equation (\ref{eq:P-3}) can be decomposed into the transmission over
the barrier (i.e. $E+\hbar\omega>\mu+\phi$) and tunneling through
the barrier as

\begin{equation}
P=P_{\mathrm{over}}+P_{\mathrm{tunnel}},
\end{equation}

\begin{equation}
P_{\mathrm{over}}=\frac{\intop_{\mu+\phi-\hbar\omega}^{\infty}T\left(\varepsilon+\hbar\omega\right)f\left(\varepsilon\right)d\varepsilon}{\intop_{0}^{\infty}f\left(\varepsilon\right)d\varepsilon}\label{eq:over}
\end{equation}

\begin{equation}
P_{\mathrm{tunnel}}=\frac{\intop_{0}^{\mu+\phi-\hbar\omega}T\left(\varepsilon+\hbar\omega\right)f\left(\varepsilon\right)d\varepsilon}{\intop_{0}^{\infty}f\left(\varepsilon\right)d\varepsilon}.\label{eq:tunnel-1}
\end{equation}

An acceptable approximation for (\ref{eq:over}) is to use the step
function as $T\left(E+\hbar\omega\right)$, as in Fowler-Dubridge model,
which leads to (\ref{eq:P-2}) and (\ref{eq:P}). More accurate expressions
can be found quantum mechanically which provides a negligible improvement. 

For electrons with energies such that $E+\hbar\omega<\mu+\phi$, the
transmission function can be found using the Gamow factor technique.
However, similar to Fowler-Nordheim formulation, we may use zero
temperature approximation for $f\left(E\right)$ and adjust the upper
integral limit as $\left(\mu+\phi-\hbar\omega\right)\rightarrow\mu$
(because there is no electron above $\mu,$ and electrons near $\mu$
have the highest probability of tunneling). Then, considering the numerator
of (\ref{eq:tunnel-1}) and (\ref{eq:FD_2}), it is straightforward
to show that 

\begin{equation}
\intop_{0}^{\mathrm{\mu}}T\left(\varepsilon+\hbar\omega\right)f\left(\varepsilon\right)d\varepsilon=\frac{2\pi\hbar}{q}J_{\mathrm{FN}}\left(\Phi\rightarrow\Phi-\hbar\omega\right)
\end{equation}
in which $J_{\mathrm{FN}}\left(\Phi\rightarrow\Phi-\hbar\omega\right)$ is
(\ref{eq:FN}) after the replacement $\Phi\rightarrow\Phi-\hbar\omega.$
Therefore, the final expression for $P$ is

\begin{equation}
P=\frac{\intop_{\mu+\phi-\hbar\omega}^{\infty}f\left(\varepsilon\right)d\varepsilon+\frac{2\pi\hbar}{q}J_{\mathrm{FN}}\left(\Phi\rightarrow\Phi-\hbar\omega\right)}{\intop_{0}^{\infty}f\left(\varepsilon\right)d\varepsilon}\label{eq:P-1}
\end{equation}

\begin{figure}
\begin{minipage}[t]{0.45\columnwidth}%
\includegraphics[width=3in]{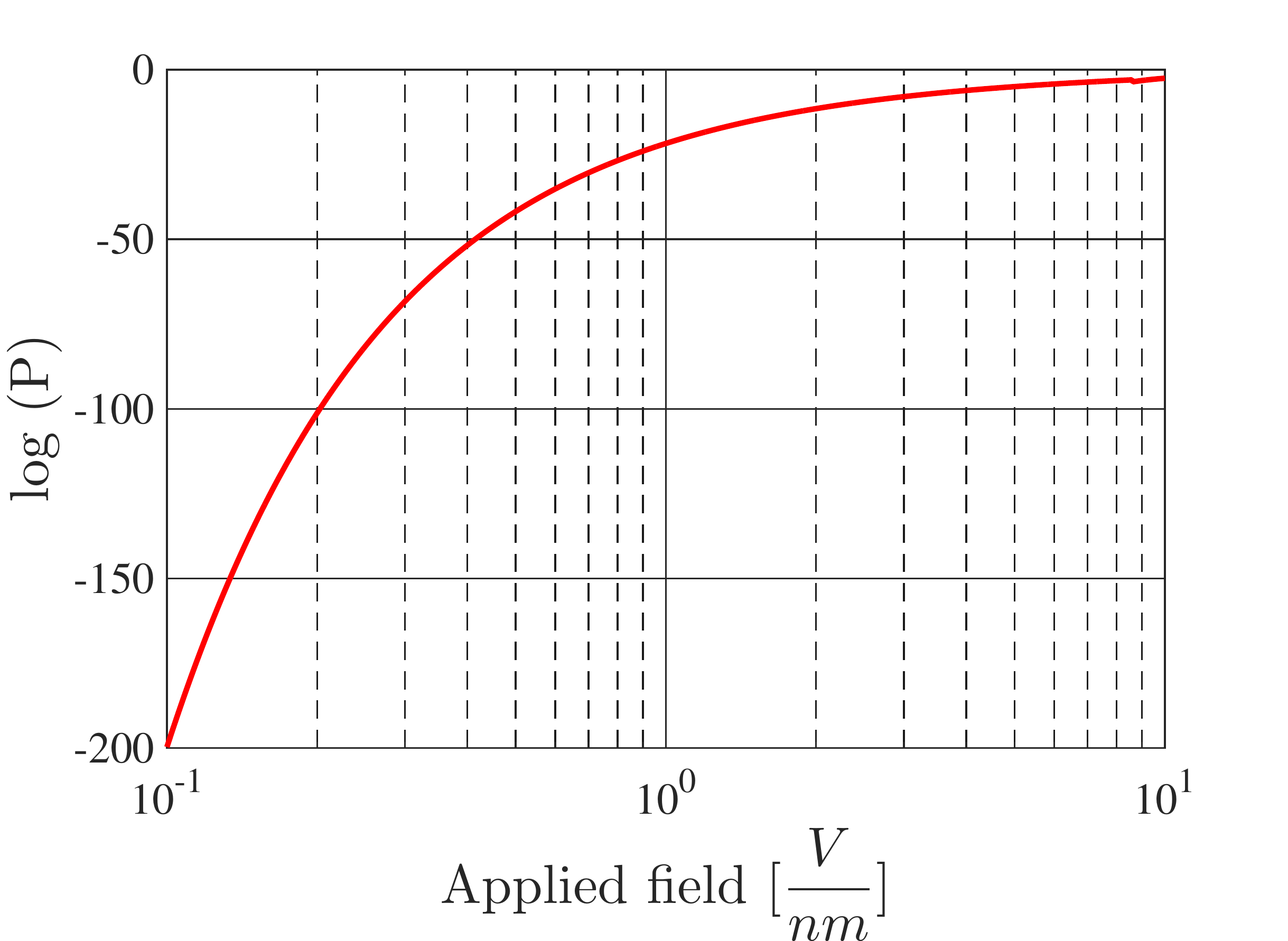}

(a)%
\end{minipage}%
\begin{minipage}[t]{0.45\columnwidth}%
\includegraphics[width=3in]{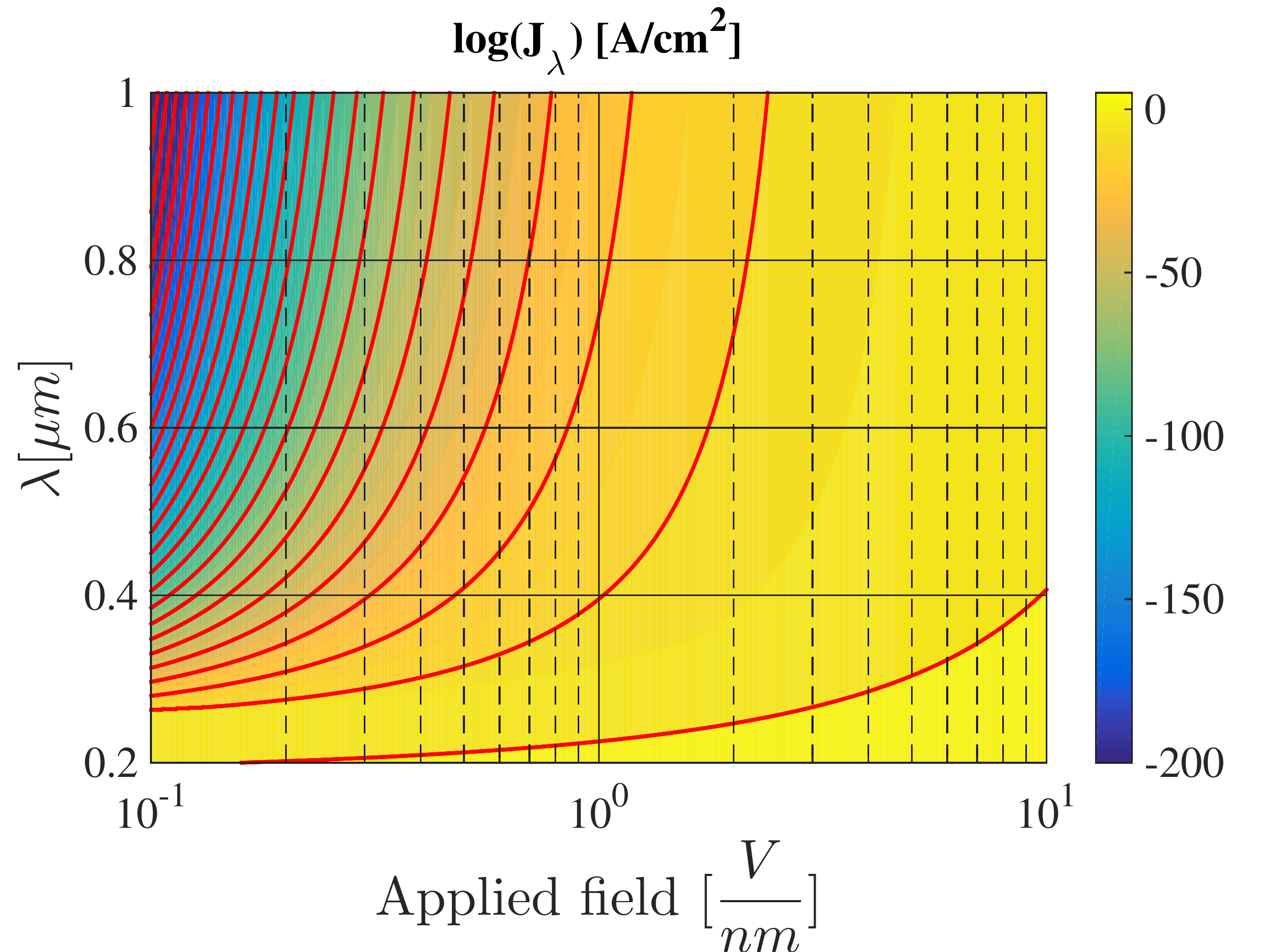}

(b)%
\end{minipage}

\caption{\label{fig:J_lambda}a) Equation (\ref{eq:P-1}) as a function of
the temperature and the applied field, b) $J_{\mathrm{\lambda}}$ at T=273$\,$K (red lines indicate isocurrent curves starting from 0 with the separation
of -10$\,\mathrm{A/cm^{2}}$)). In all calculations, gold's Fermi level of 5.53
eV and work function of 5.1 eV were used. }
\end{figure}

Equation (\ref{eq:P-1}) can be organized as 

\begin{equation}
P=\frac{U\left[\beta_{\mathrm{T}}\left(\hbar\omega-\phi\right)\right]}{U\left(\beta_{\mathrm{T}}\mu\right)}\Theta\left(\hbar\omega-\phi\right)+\frac{\pi\hbar^{2}\beta_{\mathrm{T}}^{2}F_{\mathrm{static}}^{2}}{qm\left(\Phi-\hbar\omega\right)4t^{2}U\left(\beta_{\mathrm{T}}\mu\right)}exp\left[\frac{-\nu8\pi\sqrt{2m}\left(\Phi-\hbar\omega\right)^{1.5}}{3hF}\right],\label{eq:P2}
\end{equation}
where $\Theta$ is the Heaviside step function, and indicates that
the first term should be applied only if $\hbar\omega>\phi.$ Equation
(\ref{eq:P2}), as the generalized photoemission probability, enables
us to estimate the photoemission contribution when the illuminating
laser has a photon energy lower than the lowered work function of the
cathode. Figure \ref{fig:J_lambda}(a) shows (\ref{eq:P-1}) as a
function of the applied field for a gold photocathode at $\lambda=785\,nm$ and 
T=273 K. The photoemitted current, $J_{\mathrm{\lambda}}$, as a function of
the laser wavelength and the applied field is shown in Figure \ref{fig:J_lambda}(b).
The laser intensity is assumed to be 5$\,\mathrm{W/cm^{2}}$. 

\section{Photoemission by resonant photocathodes}

If resonance effect is included in the photocathode design, it can
impact/improve all three steps of the photoemission process. An example
of such photocathodes are metallic IR photocathodes with plasmonic
resonant field enhancement \cite{putnam2017optical}. It
is common to approach resonant effects with their quality factor,
Q. However, for our purpose, it is more conveient to consider their equivalent electric field
enhancement factor  which is defined as the ratio of the electric field on the
cavity surface ($F_{\mathrm{cavity}}$) to the incident electric field $\left(F_{\mathrm{laser}}\right)$

\begin{equation}
\alpha_{\mathrm{enhance}}=\frac{F_{\mathrm{cavity}}}{F_{\mathrm{laser}}}=\frac{F_{\mathrm{cavity}}}{q\sqrt{240\pi I_{\mathrm{\lambda}}}}.
\end{equation}
Note that $F_{\mathrm{cavity}}$ for metals is normal to the cavity surface,
and it's value can be related to Q as discussed in \cite{li2017study,li2016study}.
Typical values of $\alpha_{\mathrm{enhance}}$ are in the range $10\sim100$
for noble metals at IR wavelengths. Since the electric field on the
cavity surface is enhanced by $\alpha_{\mathrm{enhance}}$, it is a good approximation
to assume that penetrated photons inside the cavity are enhanced by
a factor of $\alpha_{\mathrm{enhance}}^{2}$. The resonance effect also changes
the scattering rate inside the cavity, and the assumptions  for obtaining (\ref{eq:F_lam}) is
no longer valid (e.g. there is not a plane wave traveling inside the metal anymore). However, extracting a new $f_{\mathrm{\lambda}}$ for different
cavity shapes is beyond the scope of this paper, and we keep using
(\ref{eq:F_lam}) as an approximation. Moreover, if $\alpha_{\mathrm{enhance}}F_{\mathrm{laser}}$
becomes comparable with the applied static field, it can contribute
in the barrier reduction and the replacement $F\rightarrow\left(F_{\mathrm{static}}+F_{\mathrm{cavity}}\right)$
should be made in (\ref{eq:phi}) leading to

\begin{equation}
\phi_{\mathrm{cavity}}==\Phi-\sqrt{4Q\left(F_{\mathrm{static}}+\alpha_{\mathrm{enhance}}q\sqrt{240\pi I_{\mathrm{\lambda}}}\right)},
\end{equation}
 where $F_{\mathrm{static}}$ is the applied static field to the photocathode.
The final photoemission equation is then

\begin{equation}
J_{\mathrm{\lambda}}=I_{\mathrm{\lambda}}\left(\frac{q}{\hbar\omega}\right)\alpha_{\mathrm{enhance}}^{2}f_{\mathrm{\lambda}}P_{\mathrm{cavity}},\label{eq:J_lam-2-1}
\end{equation}
where

\[
P_{\mathrm{cavity}}=\frac{U\left[\beta_{\mathrm{T}}\left(\hbar\omega-\phi_{\mathrm{cavity}}\right)\right]}{U\left(\beta_{\mathrm{T}}\mu\right)}\Theta\left(\hbar\omega-\phi_{\mathrm{cavity}}\right)+
\]

\begin{equation}
\qquad\qquad\frac{\beta_{\mathrm{T}}}{U\left(\beta_{\mathrm{T}}\mu\right)}\frac{\left(F_{\mathrm{static}}+\alpha_{\mathrm{enhance}}q\sqrt{240\pi I_{\mathrm{\lambda}}}\right)^{2}}{\left(\Phi-\hbar\omega\right)8\pi t^{2}}exp\left[\frac{-\nu8\pi\sqrt{2m}\left(\Phi-\hbar\omega\right)^{1.5}}{3h\left(F_{\mathrm{static}}+\alpha_{\mathrm{enhance}}q\sqrt{240\pi I_{\mathrm{\lambda}}}\right)}\right],\label{eq:resonant cavity}
\end{equation}

At last, adding the thermal-field emission to the photoemission gives
the total current of the photocathode as

\begin{equation}
J=J_{\mathrm{GTF}}+J_{\mathrm{\lambda}}.
\end{equation}

\begin{figure}
\begin{minipage}[t]{0.45\columnwidth}%
\includegraphics[width=2.9in]{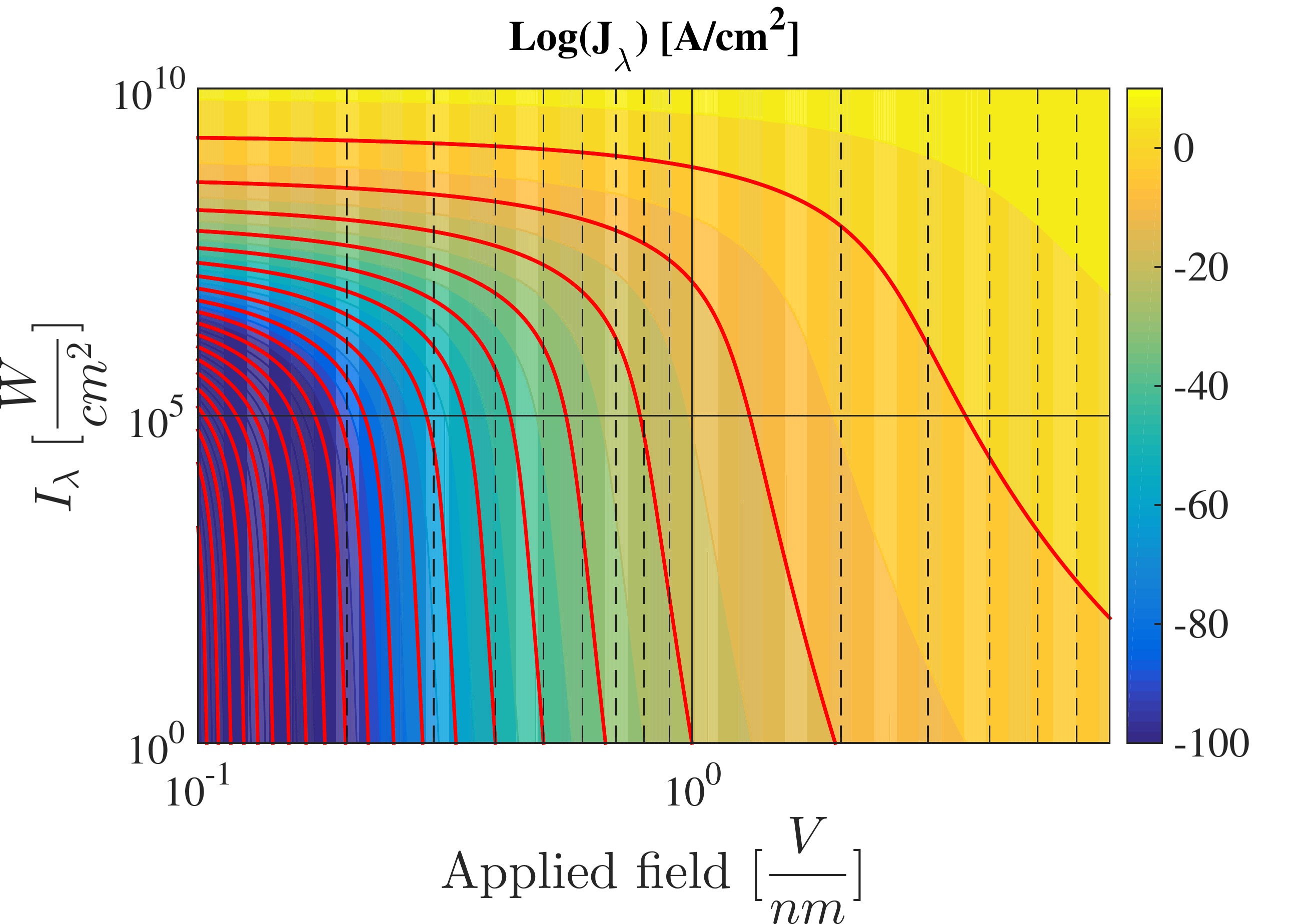}

(a)%
\end{minipage}%
\begin{minipage}[t]{0.45\columnwidth}%
\includegraphics[width=2.9in]{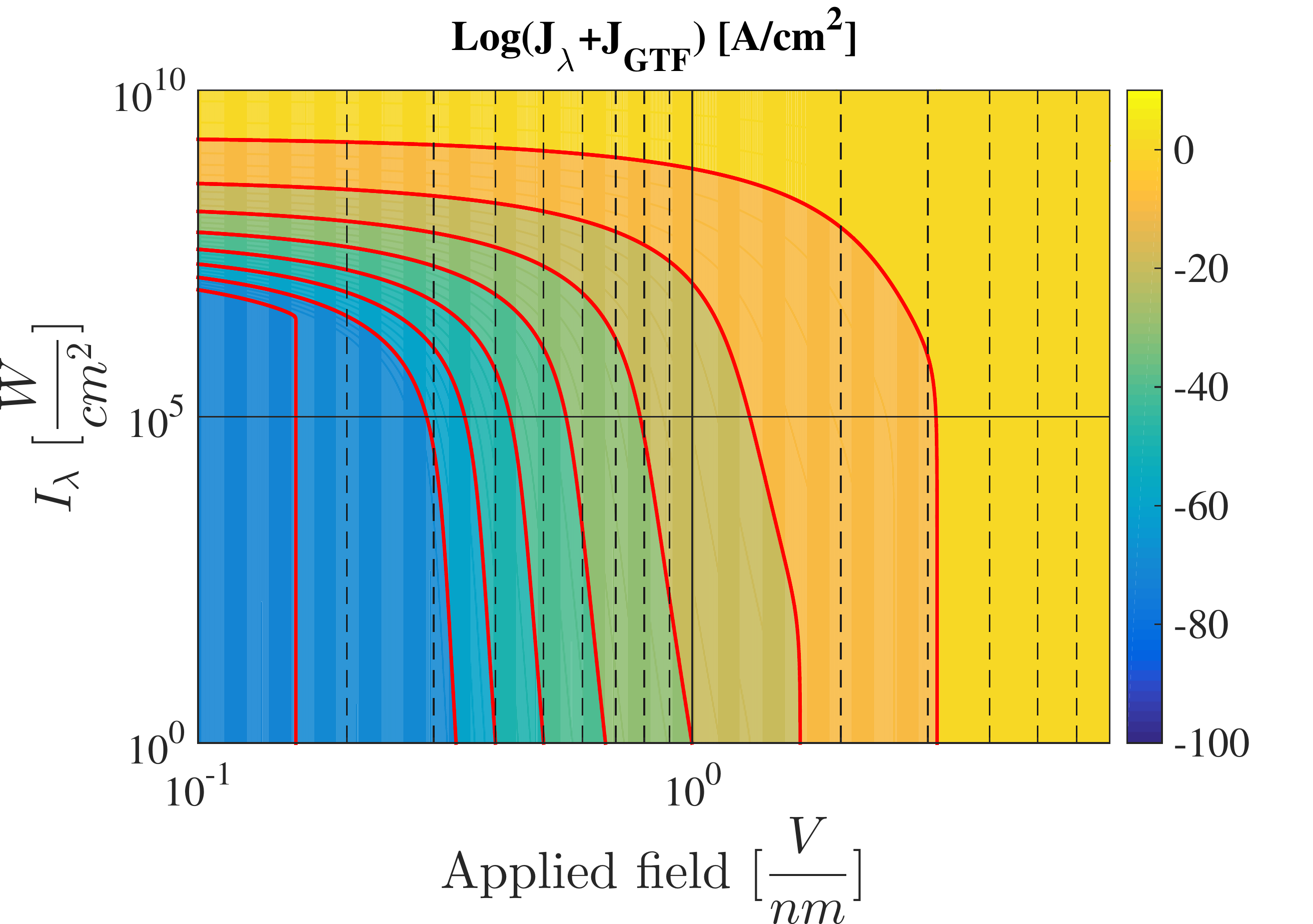}

(b)%
\end{minipage}

\caption{\label{fig:J_Lambda_res}a) $J_{\mathrm{\lambda}}$ and b) $J_{\mathrm{\lambda}}+J_{\mathrm{GTF}}$
of a resonant photocathode with $\alpha_{\mathrm{enhance}}=20$ at T=273$\,$K
(red lines indicate isocurrent curves starting from 0 with the separation
of -10$\,\mathrm{A/cm^{2}}$)). In all calculations, gold's Fermi level of 5.53$\,$eV and work function of 5.1$\,$eV were used. }
\end{figure}

\begin{figure}
\includegraphics[width=3in]{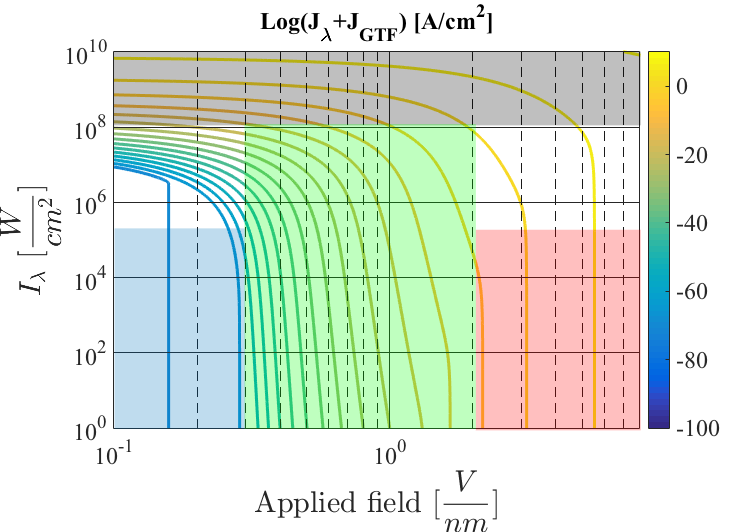}\caption{\label{fig:J_total_regions} Different emission regions of the total
current (explanation in the text). Parameters are the same as Fig.
\ref{fig:J_Lambda_res}. }
\end{figure}

Figure \ref{fig:J_Lambda_res} shows the photo-emitted current component,
$J_{\mathrm{\lambda}}$, and the total current, $J_{\mathrm{\lambda}}+J_{\mathrm{GTF}}$, as
a function of the irradiance power and the applied field. The optical
enhancement factor of 20 is assumed for the photocathode, the wavelength
is fixed at 785 nm, and gold parameters are used for the photocathode.
Four important regions of the total current are specified in Fig.
\ref{fig:J_total_regions}. The blue region is where both field emission
and photoemission currents are negligible, and only thermal emission
at 273 K (a small number) exists. In the red region, field emission
is dominant and changing the laser intensity does not have much effect
on the total current. Inversely, in the gray region, photoemission
current dominates the total current and the applied static field does
not have a comparable effect. The green region is where both field
emission and photoemission are comparable, and is useful for designing
microelectronic devices with both voltage and optical controls. This
region is identified by $0.2<F_{\mathrm{static}}[\mathrm{\frac{V}{nm}}]<2$
and $I_{\mathrm{\lambda}}<100\,\mathrm{\frac{MW}{cm^{2}}}$ for gold photocathode
at $\lambda=785\,$nm, and can be obtained for different materials
in a similar fashion. As the final note on this section, a very useful
list of experimental parameters for different metals are given in
Table I of \cite{jensen2006photoemission}.

\begin{figure}
\begin{minipage}[t]{0.4\columnwidth}%
\includegraphics[width=2.7in]{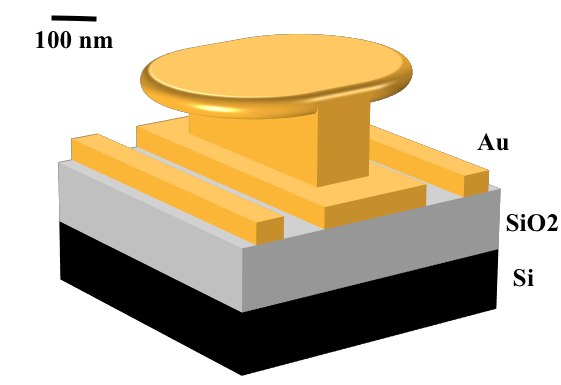}

(a)%
\end{minipage}%
\begin{minipage}[t]{0.4\columnwidth}%
\includegraphics[width=3in]{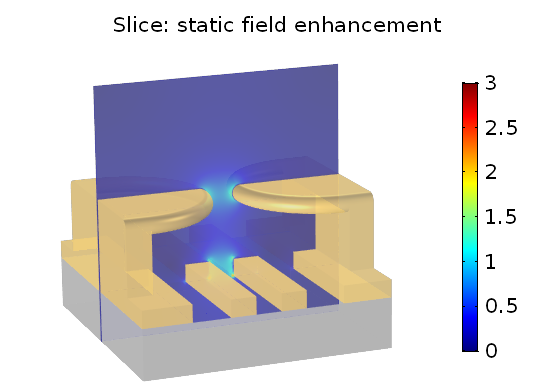}

(b)%
\end{minipage}

\caption{\label{fig:Unit_cell}a) unit cell of the resonant photocathode (array
of 12 by 12 elements), b) simulated static electric field enhancement
of the photocathode. The smallest gap size is 100 nm, and the metal
is gold (see \cite{forati2016photoemission,forati2016fabrication}).}
\end{figure}

\begin{figure}
\includegraphics[width=3in]{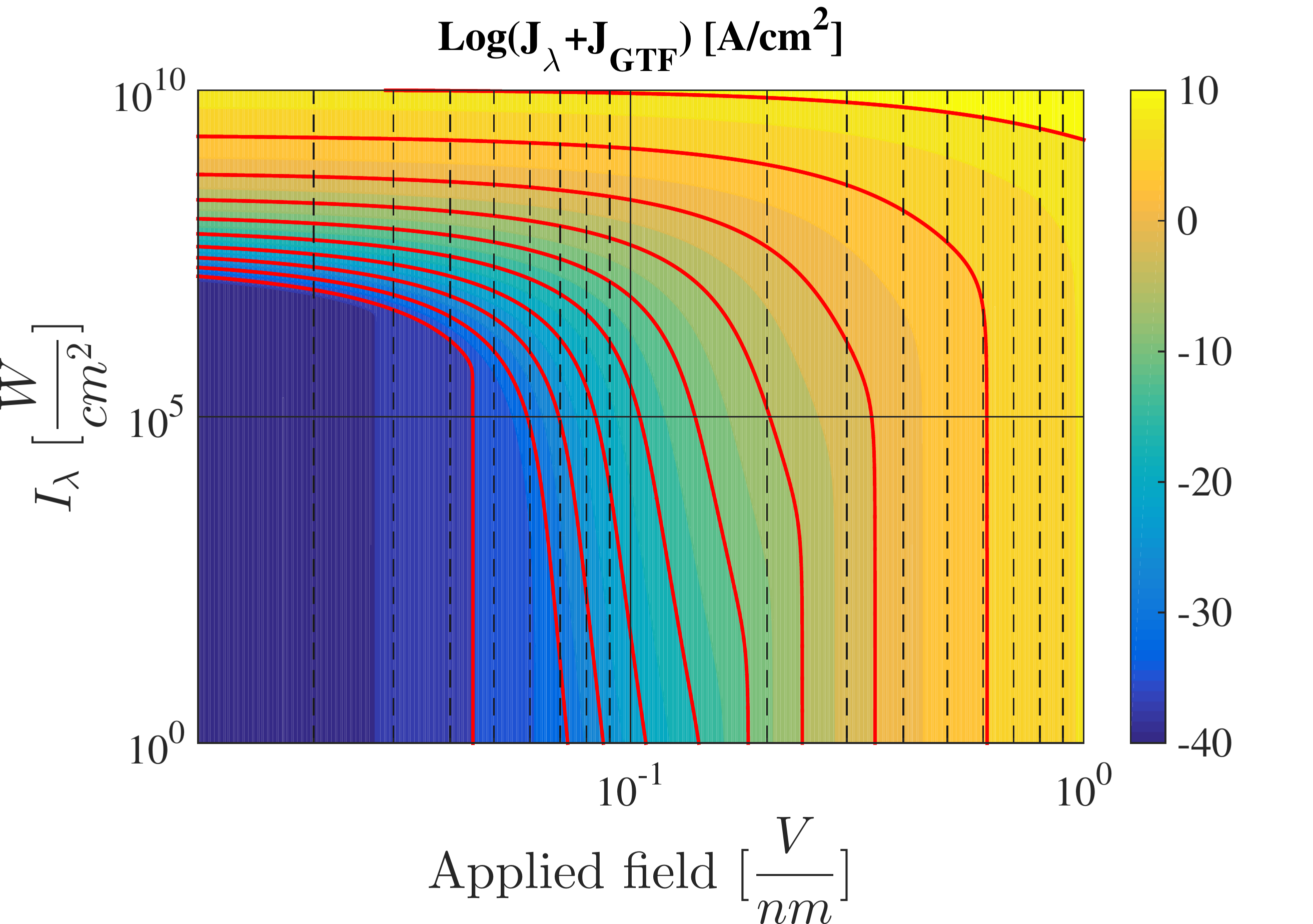}\caption{\label{fig:J_total} Total current ($J_{\mathrm{\lambda}}+J_{\mathrm{GTF}}$) of the
photocathode at T=550 K with the optical and static enhancement factors
of 37.2 and 9, respectively. Red lines indicate isocurrent curves
starting from 10 with steps of -5$\,\mathrm{A/cm^{2}}$. Gold's Fermi
level of 5.53$\,$eV and work function of 5.1$\,$eV were used. }

\end{figure}

\section{Example}

Figure \ref{fig:Unit_cell}(a) shows the unit cell of the IR photocathode
studied in \cite{forati2016photoemission}, experimentally. The photocathode
is fabricated using gold on Si substrate with a $200\,$nm $SiO_{2}$
in between as isolation. The surface area of the photocathode is around
450 $\mu\,$m (22 by 22 unit cells with the area of about 1 by 1 micron),
and the minimum gap size is $100\,$nm. As reported in \cite{forati2016photoemission},
full wave simulated (using COMSOL) and measured optical electric field
enhancement of the photocathode were 12 and 37, respectively. The
factor of 3 difference was due to the surface roughness of the fabricated
device, which was not considered in the simulations. Solving Poisson's
equation on the same unit cell in COMSOL leads to the static electric
field enhancement of $2.5\sim3$, as Fig. \ref{fig:Unit_cell}(b)
shows. Using the additional factor of 3 due the surface roughness,
we perform our calculations with the static electric filed enhancement
of 9. Figure \ref{fig:J_total} shows the total current emission of
the photocathode as a function of the applied static field and the laser
intensity. The wavelength is 875 nm, the temperature is assumed to
be 550 K (which seems to be a reasonable guess). Complex permittivity
values of $\epsilon_{\mathrm{Si}}=13.734-0.054952i$ and $\epsilon_{\mathrm{SiO_{2}}}=2.1502-0i$
were used for silicon and $SiO_{2},$ respectively \cite{Refractive_index}. 

\begin{figure}
\includegraphics[width=3.5in]{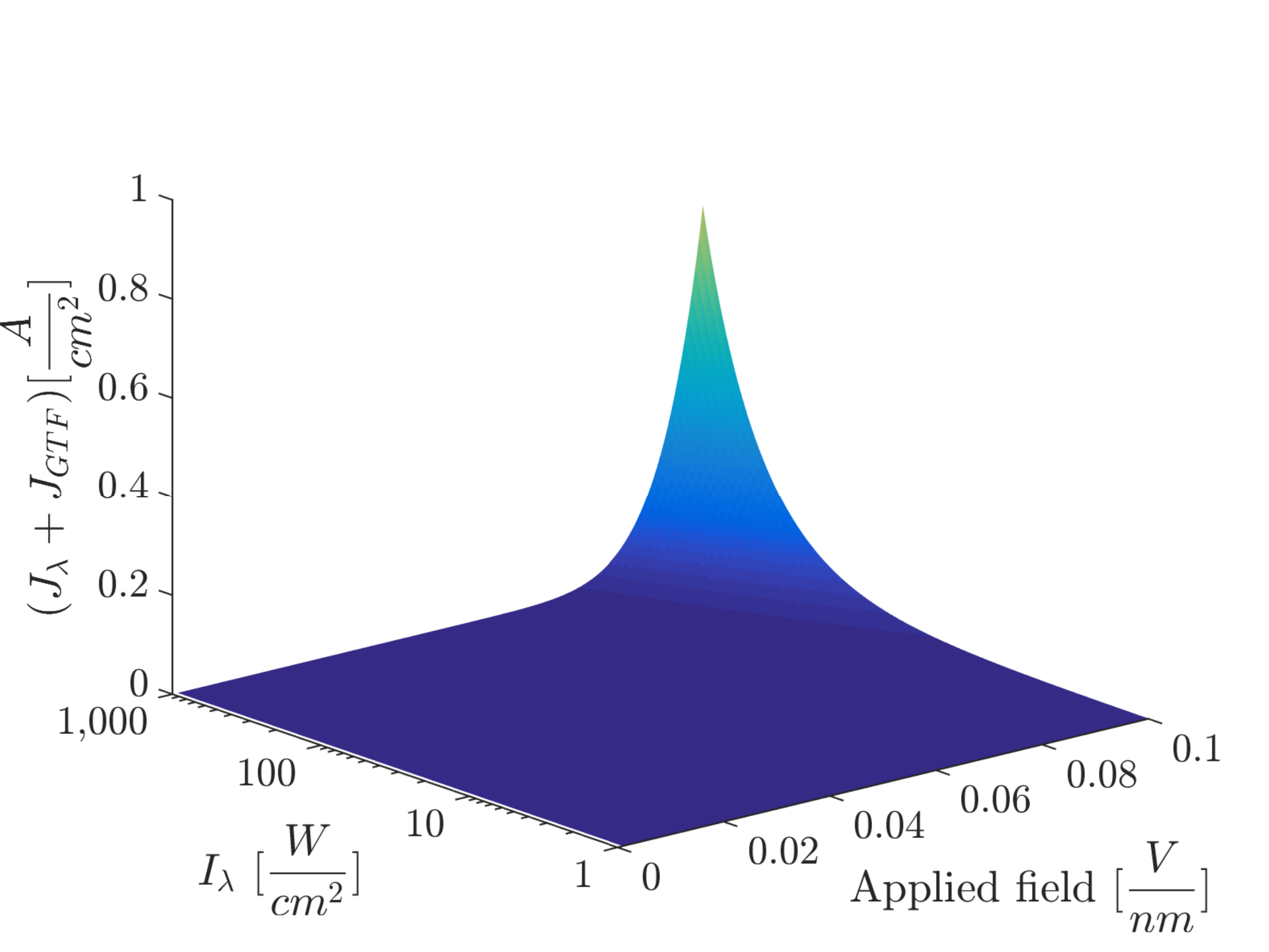}\caption{\label{fig:J_total_Linear} Total current ($J_{\mathrm{\lambda}}+J_{\mathrm{GTF}}$) of the resonant
photocathode at T=550 K with the optical and static enhancement factors
of 300 and 10, respectively. Gold's Fermi
level of 5.53$\,$eV and work function of 5.1$\,$eV were used. }

\end{figure}

Comparing the the experimental values reported in \cite{forati2016photoemission}
with Fig. \ref{fig:J_total}, either a higher field enhancement
factor (for the static or the laser field) should be considered, or the thermal
effect is considerable. Note that our suggested theory for long wavelength
photoemission, including (\ref{eq:resonant cavity}), was based on a
low temperature assumption. The approximations in our formulations
may not be valid if the laser intensity can change the photocathode
temperature considerably. With the available experimental data reported
in \cite{forati2016photoemission}, we cannot confirm if the temperature of the photocathode
rises significantly. But, plasmoinc resonance effects are claimed to be capable of providing laser field enhancements on the order of 1000 \cite{ward2010optical}. An important conclusion can be made if draw Fig.  \ref{fig:J_total} on a linear scale, as shown in Fig. \ref{fig:J_total_Linear} (note that field enhancements are changed to 10 and 1000 for the static and laser fields, respectively). Figure \ref{fig:J_total}  shows that, above a certain threshold, simultaneous application of both laser and static fields can cause significant electron emission. In contrast, neither static nor optical field alone generates nearly as much current as both fields combined. This result is also consistent with the results in \cite{piltan2017field} which are based on the analytical solution for Schrodinger equation with a triangular work function \cite{zhang2016ultrafast}. The analytical solution in \cite{zhang2016ultrafast} is general and considers the multiphoton absorption/emission effect. However, the results in this paper are only based on single photon absorption, hence their accuracy becomes questionable at strong laser fields (intensities above 1e10 ${\mathrm{(W/cm^2}}$). Note that the results in \cite{zhang2016ultrafast} unnecessarily disregards the charge image effect (Schottkey effect) in most of the generated results (which is significant for the applied static field range). Moreover, the solution in \cite{zhang2016ultrafast} only assumes electrons at the Fermi level while the results in this paper includes the supply function of the metal.

\section{Conclusion}

Electron emission theories were reviewed and a formulation was extracted
for the electron emission by photocathodes under exposure of long
wavelength lasers.  With the summarized formulations in this paper,
an optimal range for the applied static field and laser illumination
was obtained for gold photocathodes so that both $J_{\mathrm{\lambda}}$ and
$J_{\mathrm{F}}$ are comparable. This condition is needed for the design of
nano-electronic devices with both optical and voltage controls. The
effect of optical resonance was also shown to be significant in the
performance of photocathodes. This can include generation of electron
beams with strong correlation with the incoming photons.

\section{Acknowledgement}

The authors would like to thank Dr. Kevin Jensen for his generous
help to better understand his theories and results. 

\bibliographystyle{apsrev4-1}
\bibliography{Electron_emission}

\section{Appendix}

\subsection{The three step model of Spicer and Perrera-Gomez \cite{spicer1993modern}}

The three steps of the model are a) optical absorption by the material,
b) electron transport inside the material, and c) electron emission
(escape) from the material surface. The laser intensity inside the
material can be written as

\begin{equation}
S\left(x,\lambda\right)=S_{0}\left(1-R\right)e^{-\alpha x}
\end{equation}

where $S_{0}$ is the incident laser intensity, $R$ is the reflection
from the material surface which depends on the wavelength, $\alpha$
is the laser attenuation inside the material both due to absorption
by electrons and phonons, and $x$ is the 1D coordinate with the material
boundary at $x=0$. Note that both $R$ and $\alpha$ are wavelength
dependent.

The light absorption is the rate of the light intensity decrease as 

\begin{equation}
dS\left(x,\lambda\right)=\alpha S_{0}\left(1-R\right)e^{\mathrm{-\alpha x}}dx.
\end{equation}
Consider a slab between $x$ and $x+dx$. Only a portion of the absorbed
photons in this slab can lead to electron liberation from the material
surface. The contribution of these photons into the emitted current
can be written as 

\begin{equation}
di\left(x\right)=P_{o\alpha}\left(\lambda,x,dx\right)P_{\mathrm{T}}\left(\lambda,x\right)P_{\mathrm{E}}\left(\lambda\right)\label{eq:i_x}
\end{equation}
where $P_{o\alpha}$ is the probability of exciting electrons above
the vacuum level (VL) by photons absorbed in the slab between x and
x+dx, $P_{\mathrm{T}}$ is the probability that electrons will reach the surface
without losing their sufficient energy, and $P_{\mathrm{E}}$ is the probability
of escape from the surface. Let us define $\alpha_{\mathrm{PE}}$ as the probability
of exciting an electron above the VL by a photon so that

\begin{equation}
P_{o\alpha}\left(x,dx\right)=\alpha_{\mathrm{PE}}I_{x}dx=\alpha_{\mathrm{PE}}I_{0}\left(1-R\right)e^{\mathrm{-\alpha x}}dx,
\end{equation}
and define the escape length, $L$, so that

\begin{equation}
P_{\mathrm{T}}=exp\left(-\frac{x}{L}\right).
\end{equation}

Then, (\ref{eq:i_x}) can be written as 

\begin{equation}
di\left(x\right)=\alpha_{\mathrm{PE}}I_{0}\left(1-R\right)e^{\mathrm{-\alpha x}}e^{\mathrm{-x/L}}P_{\mathrm{E}}dx.
\end{equation}

This gives us QE as 

\begin{equation}
QE=\frac{1}{I_{0}}\intop_{0}^{\infty}di\left(x\right)dx=\left(1-R\right)e^{\mathrm{-\alpha x}}\frac{\alpha_{\mathrm{PE}}P_{\mathrm{E}}}{\alpha+1/L}.\label{eq:QE}
\end{equation}

It is usual to define an absorption length as $l_{\mathrm{\alpha}}=1/\alpha$
and re-write \ref{eq:QE} as 

\begin{equation}
QE=\left(1-R\right)e^{\mathrm{-\alpha x}}\frac{\frac{\alpha_{\mathrm{PE}}}{\alpha}P_{\mathrm{E}}}{1+l_{\mathrm{\alpha}}/L}.\label{eq:QE2}
\end{equation}
 Parameters $L,$ $P_{\mathrm{E}},$ and $\alpha_{\mathrm{PE}}$ are measured experimentally
and reported for different photocathodes. Also, from Maxwell's equations,
$R$ and $\alpha$ can be calculated knowing the permittivity of the
photocathode material. 

The ratio $\alpha_{\mathrm{PE}}/\alpha$ is the fraction of electrons excited
above the VL. This ratio is maximized by lowering the vacuum level
to be close to the conduction band minimum (CBM). This ratio for effective
emitters is usually between 0.1 and 1. Emitters with negative electron
affinity have this ratio at around unity. The negative affinity cathodes
were the first ``significantly engineered'' photocathodes introduced
in 1965. The parameter $\alpha_{\mathrm{PE}}$ usually increases with the photon
energy. 

By examining the ratio $l_{\mathrm{a}}/L$, it is clear that faster absorption
of the light leads to a higher QE. In other words, high energy photons
are disadvantageous in this regard since they can travel far into
the metal. The parameter $P_{\mathrm{E}}$ is usually less than 0.1 for metals,
and it usually increases monolithically with the incident photon energy. 

\subsection{The general field-thermal model with temperature coupling (short
wavelength)}

This section is a summary of the model presented in \cite{jensen2007general},
which is the most comprehensive field-thermal emission model so far.
This model calculates the electron and lattice temperatures using
coupled differential equations which relates temperature to the laser
power. Then, using thermal conductivity and optical parameters of
materials, scattering and reflection of electrons and photons are
calculated. The key point in this model is to find temperatures of
electrons and lattice due to laser illumination. The following steps
should be considered based on this model:

1) Find the absorbed laser power from 

\begin{equation}
G\left(z,t\right)=I_{\mathrm{\lambda}}\left(t\right)\left(1-R\right)\frac{e^{\mathrm{-z/\delta}}}{\delta}\left\{ 1-P\left(\hbar\omega\right)\right\} 
\end{equation}

in which $R$ is the reflection from the material and $\delta=\frac{\lambda}{4\pi k}$
is the skin depth (note that $n+ik$ is known for the emitter material).
For a two dimensional material, this parameter can be extracted from
Snell's law (for parallel and transverse polarizations), leading to

\begin{equation}
R=0.5\left(R_{\mathrm{s}}+R_{\mathrm{p}}\right),
\end{equation}
\begin{equation}
R_{\mathrm{s}}=\frac{a^{2}+b^{2}-2acos\theta+cos^{2}\theta}{a^{2}+b^{2}+2acos\theta+cos^{2}\theta},
\end{equation}

\begin{equation}
R_{\mathrm{p}}=R_{\mathrm{s}}\frac{a^{2}+b^{2}-2asin\theta tan\theta+sin^{2}\theta tan^{2}\theta}{a^{2}+b^{2}+2asin\theta tan\theta+sin^{2}\theta tan^{2}\theta},
\end{equation}

\begin{equation}
2a^{2}=\left[\left(n^{2}-k^{2}-sin^{2}\theta\right)^{2}+\left(2nk\right)^{2}\right]^{0.5}+\left(n^{2}-k^{2}-sin^{2}\theta\right),
\end{equation}

\begin{equation}
2b^{2}=\left[\left(n^{2}-k^{2}-sin^{2}\theta\right)^{2}+\left(2nk\right)^{2}\right]^{0.5}-\left(n^{2}-k^{2}-sin^{2}\theta\right).
\end{equation}

$\theta$ is the incidence angle. 

2) Find electron and lattice specific heat parameters as

\begin{equation}
C_{\mathrm{e}}\left(T_{\mathrm{e}}\right)=\frac{\gamma T_{\mathrm{e}}}{1+7/40\left(\pi/\beta_{\mathrm{e}}\mu\right)^{2}}
\end{equation}

\begin{equation}
C_{\mathrm{i}}\left(T_{\mathrm{i}}\right)=\frac{3Nk_{\mathrm{B}}}{1+1/20\left(T_{\mathrm{D}}/T_{\mathrm{i}}\right)^{2}}
\end{equation}

where $\beta_{\mathrm{e}}=1/k_{\mathrm{B}}T_{\mathrm{e}}$, and $N\left(\#/cm^{3}\right)$ is
the number density of the crystal. Deby temperature and $\gamma$
are given in table I of \cite{jensen2007general}. 

3) Find the scattering factor $f_{\mathrm{\lambda}}$ which accounts for the
probability of an electron traveling to the surface of a metal, without
scattering, with a kinetic energy component normal to the surface,
sufficient for emission. The scattering factor can be found from 

\begin{equation}
f_{\mathrm{\lambda}}=0.5G\left(z_{0}\right)+0.5z_{0}G^{\prime}\left(z_{0}\right)\left[\frac{3k_{\mathrm{v}}P-\left(k_{\mathrm{F}}-k_{o}\right)^{2}\left(2k_{\mathrm{F}}+k_{0}\right)}{3\left(2k_{\mathrm{F}}+k_{0}\right)\left(k_{\mathrm{F}}-k_{0}\right)^{2}}\right]\label{eq:f_lambda-1}
\end{equation}

\begin{equation}
P=0.5\left(k_{\mathrm{\lambda}}^{2}+2k_{\mathrm{F}}^{2}\right)ln\left(\frac{k_{\mathrm{F}}+k_{\mathrm{u}}}{k_{0}+k_{\mathrm{v}}}\right)-0.5\left(k_{\mathrm{F}}k_{\mathrm{u}}-k_{0}k_{\mathrm{v}}\right)
\end{equation}

\begin{equation}
z_{0}=\frac{\delta m}{\hbar k_{\mathrm{v}}\tau}
\end{equation}

\begin{equation}
\left(\hbar k_{\mathrm{u}}\right)^{2}=2m\left(\mu+\hbar\omega\right)
\end{equation}

\begin{equation}
\left(\hbar k_{\mathrm{v}}\right)^{2}=2m\left(\mu+\phi\right)
\end{equation}

\begin{equation}
\left(\hbar k_{0}\right)^{2}=2m\left(\mu+\phi-\hbar\omega\right)
\end{equation}

\begin{equation}
\left(\hbar k_{\mathrm{F}}\right)^{2}=2m\mu
\end{equation}

\begin{equation}
G\left(x\right)=1-\frac{2arccos\left(1/x\right)}{\pi sin\left(arccos\left(1/x\right)\right)}
\end{equation}

Note that $\tau$ is temperature dependent. Also, a good approximation
is

\begin{equation}
f_{\mathrm{\lambda}}=0.5G\left(z_{0}\right)
\end{equation}

4- Find thermal conductivity

\begin{equation}
\kappa\left(T_{\mathrm{e}},T_{i}\right)=\frac{2\mu}{3m}\tau\left(T_{\mathrm{e}},T_{\mathrm{i}}\right)C_{\mathrm{e}}\left(T_{\mathrm{e}}\right)
\end{equation}

where $\tau$ is the relaxation time as

\begin{equation}
\frac{1}{\tau}=\frac{1}{\tau_{\mathrm{ee}}\left(T_{\mathrm{e}}\right)}+\frac{1}{\tau_{\mathrm{ep}}\left(T_{\mathrm{i}}\right)}
\end{equation}

\begin{equation}
\tau_{\mathrm{ee}}\left(T_{\mathrm{e}}\right)=\frac{\hbar\mu}{A_{0}}\left(\frac{1}{k_{\mathrm{B}}T_{\mathrm{e}}}\right)^{2}=\frac{A_{\mathrm{ee}}}{T_{\mathrm{e}}^{2}}
\end{equation}

\begin{equation}
\tau_{\mathrm{ep}}\left(T_{\mathrm{i}}\right)=\frac{\hbar}{2\pi\lambda_{0}}\left(\frac{1}{k_{\mathrm{B}}T_{\mathrm{i}}}\right)=\frac{B_{\mathrm{ep}}}{T_{\mathrm{i}}}
\end{equation}

Parameters $\lambda_{0}$ and $A_{0}$ should be extracted from Fig.
4 of \cite{jensen2007general}

5- Use the obtained $\lambda_{0}$ to find the g parameter as

\begin{equation}
g=\frac{\pi\lambda_{0}}{9\hbar}k_{\mathrm{B}}k_{\mathrm{F}}^{3}m\nu_{\mathrm{s}}^{2}
\end{equation}

which governs the transfer of energy from electron to the lattice.

Note that at steady state, $T_{\mathrm{e}}=T_{\mathrm{i}}$. For pulse widths longer
than nanoseconds, since the scattering rates are much smaller, this
steady state approximation is valid and we may use
\begin{equation}
T_{\mathrm{e}}\left(z\right)=T_{0}+\left(T_{\mathrm{e}}-T_{0}\right)exp\left(-\frac{zm}{n\hbar k_{\mathrm{F}}\tau}\right)
\end{equation}

where the factor n is on the order of the square root of the ratio
of the pulse time to the scattering time: $n=\left(1ns/0.1ps\right)^{0.5}.$

6- Find the temperature distribution by solving coupled differential
equations \cite{papadogiannis1997electron,papadogiannis2001ultrashort}

\begin{equation}
C_{\mathrm{e}}\left(T_{\mathrm{e}}\right)\frac{\partial}{\partial t}T_{\mathrm{e}}=\frac{\partial}{\partial z}\left[\kappa\left(T_{\mathrm{e}},T_{\mathrm{i}}\right)\frac{\partial}{\partial z}T_{\mathrm{e}}\right]-g\times\left(T_{\mathrm{e}}-T_{\mathrm{i}}\right)+G\left(z,t\right)
\end{equation}

\begin{equation}
C_{\mathrm{i}}\left(T_{\mathrm{i}}\right)\frac{\partial}{\partial t}T_{\mathrm{i}}=g\times\left(T_{\mathrm{e}}-T_{\mathrm{i}}\right)
\end{equation}

Note that the surface is located at z=0 and negative z indicates inside
of the metal. 

Solution of the above coupled differential equations also provides
the required parameters for calculating QE accurately (considering
the heat effect).

As a final point, note that electrons can be excited to energies above
the potential barrier inside the metal. However, a static field is
always needed to bring them to the surface. This cannot be done by
the laser electric field because of its small penetration depth, and
it's small wavelength inside the metal (which only can cause electrons
oscillations, with zero net displacement).

\end{document}